\documentclass[12pt]{article}
\pdfoutput=1
\usepackage{pstricks}
\usepackage{color}
\usepackage{cite}
\usepackage{array}
\usepackage{epsfig}
\usepackage{amssymb}
\usepackage{graphics,graphpap}
\usepackage{amssymb}
\usepackage{amsmath}
\usepackage{slashed}
\usepackage{dsfont}

\newcommand{\RE}{{\rm Re}}
\newcommand{\IM}{{\rm Im}}

\newcommand{\tev}{\, {\rm TeV}}
\newcommand{\gev}{\, {\rm GeV}}
\newcommand{\mev}{\, {\rm MeV}}

\newcommand{\be}{\begin{equation}}
\newcommand{\ee}{\end{equation}}
\newcommand{\bea}{\begin{eqnarray}}
\newcommand{\eea}{\end{eqnarray}}

\newcommand{\bi}{\begin{itemize}}
\newcommand{\ei}{\end{itemize}}
\newcommand{\ord}{{\cal O}}

\newcommand{\vcb}{|V_{cb}|}
\newcommand{\vtd}{|V_{td}|}
\newcommand{\vub}{|V_{ub}|}
\newcommand{\vts}{|V_{ts}|}

\usepackage{graphicx}

\medskipamount 

\usepackage{fancyhdr}
\advance \headheight by 3.0truept       

\newlength{\textlength}
\newlength{\overlinelength}

 \def\s#1{\setbox0=\hbox{$#1$}%
   \rlap{\ifdim\wd0>.7em\kern.22\wd0\else\kern.1\wd0\fi /}#1}


\begin{document}

\begin{titlepage}
\begin{flushright}
{FLAVOUR(267104)-ERC-50}
\end{flushright}
\vskip1.0cm
\begin{center}
{\Large \bf \boldmath
Left-handed $Z^\prime$ and $Z$ FCNC quark couplings facing new $b\to s\mu^+\mu^-$ 
data}
\vskip1.0cm
{\bf
Andrzej J. Buras and 
Jennifer Girrbach}
\vskip0.3cm
TUM-IAS, Lichtenbergstr. 2a, D-85748 Garching, Germany\\
Physik Department, TUM, D-85748 Garching, Germany\\

\vskip0.41cm


\vskip0.35cm

{\large\bf Abstract\\[10pt]} \parbox[t]{\textwidth}{
In view of the recent improved data on $B_{s,d}\to\mu^+\mu^-$ and 
$B_d\to K^*\mu^+\mu^-$ we revisit two simple New Physics (NP) scenarios 
analyzed by us last year in which new FCNC currents in $b\to s \mu^+\mu^-$ 
transitions are mediated either
entirely by a neutral heavy gauge boson $Z^\prime$ with purely 
left-handed complex couplings $\Delta_L^{qb}(Z^\prime)$ ($q=d,s$) and real 
couplings to muons $\Delta_A^{\mu\bar\mu}(Z^\prime)$ and 
$\Delta_V^{\mu\bar\mu}(Z^\prime)$ or the SM $Z$ boson with left-handed 
complex couplings $\Delta_L^{qb}(Z)$. We demonstrate how the 
{\it reduced couplings}, the couplings in question divided by $M_{Z^\prime}$ or 
$M_Z$,  can be determined by future  $\Delta F=2$ and $b\to s\mu^+\mu^-$ 
observables up to sign ambiguities. The latter do not affect the correlations between various observables that can test these NP scenarios. We present the results 
as functions of  $C_{B_q}=\Delta M_{q}/(\Delta M_{q})_{\rm SM}$, 
$S_{\psi\phi}$ and $S_{\psi K_S}$ which should be precisely determined in  this decade.  We calculate the violation of the CMFV relation
between 
${\mathcal{B}}(B_{s,d}\to\mu^+\mu^-)$ and $\Delta M_{s,d}$  in these scenarios.
We find that the data on $B_{s,d}\to\mu^+\mu^-$ from CMS and LHCb can 
be reproduced in both scenarios but in the case of $Z$, $\Delta M_s$ and 
$S_{\psi\phi}$ have to be very close to their SM values. As far as $B_d\to K^*\mu^+\mu^-$ anomalies 
are concerned the $Z^\prime$ scenario can significantly 
soften these anomalies while 
the $Z$ boson fails badly because of the small vector coupling to muons.
We also point out
that recent proposals of explaining these
anomalies with the help of a real Wilson coefficient $C^{\rm NP}_9$ implies uniquely an {\it enhancement} of $\Delta M_s$ with respect  
to its SM value,  while a complex $C^{\rm NP}_9$ allows for both enhancement and suppression of $\Delta M_s$ and simultaneously novel
CP-violating effects. 
Correlations between $b\to s\mu^+\mu^-$ and $b\to s\nu\bar\nu$ observables in these scenarios are emphasized. 
We also discuss briefly scenarios in which the $Z^\prime$ boson
has right-handed FCNC couplings.   In this context we  point out 
a number of correlations between angular observables measured in $B_d\to K^*\mu^+\mu^-$ that arise in the absence of new CP-violating phases in scenarios with only left-handed or right-handed couplings or scenarios in which 
left-handed and right-handed couplings are equal to each other or differ by sign.}

\vfill
\end{center}
\end{titlepage}

\setcounter{footnote}{0}

\newpage
\tableofcontents

\section{Introduction}
\label{sec:1}
The correlations between flavour observables in concrete New Physics (NP) models 
are a powerful tool to distinguish between various models and to select the 
ones that are consistent with the data \cite{Buras:2013ooa}. Among prominent examples 
where such correlations are rather stringent are models with constrained minimal 
flavour violation (CMFV) \cite{Buras:2000dm,Buras:2003jf}, MFV at large
\cite{Chivukula:1987py,Hall:1990ac,D'Ambrosio:2002ex},  GMFV \cite{Kagan:2009bn}, models with 
$U(2)^3$ flavour symmetry \cite{Barbieri:2011ci,Barbieri:2012uh,Crivellin:2011fb,Buras:2012sd} and  supersymmetric models  with flavour
symmetries \cite{Altmannshofer:2009ne}.

 Also  models in which all FCNCs are mediated
entirely by a neutral have gauge boson $Z^\prime$ imply a multitude of correlations as analyzed in detail 
in \cite{Buras:2012dp,Buras:2012jb,Buras:2013td,Buras:2013uqa,Buras:2013rqa}. 
 A review of $Z^\prime$ models can be found in  \cite{Langacker:2008yv} and 
other recent studies in these models have been presented in 
\cite{Barger:2009qs,Fox:2011qd,Altmannshofer:2011gn,Altmannshofer:2012ir,Dighe:2012df,Sun:2013cza}.

 While FCNC $Z^\prime$ 
couplings to quarks could be generally left-handed and right-handed, as demonstrated in particular in \cite{Buras:2012jb}, a very interesting scenario is the LHS one in which $Z^\prime$ couplings to quarks are  purely 
left-handed. The nice virtue of this scenario is that for certain choices 
of the $Z^\prime$ couplings the model resembles the structure of CMFV or models with $U(2)^3$ flavour symmetry. Moreover as no new operators beyond those 
present in the SM are present, the non-perturbative uncertainties are the 
same as in the SM, still allowing for non-MFV contributions beyond those present in $U(2)^3$ models. In particular the stringent CMFV relation between 
$\Delta M_{s,d}$ and $\mathcal{B}(B_{s,d}\to\mu^+\mu^-)$ 
\cite{Buras:2003td} valid in the simplest $U(2)^3$ models is violated 
in the LHS scenario as we will see below.

Another virtue of the LHS scenario is the paucity of its parameters 
that enter all flavour observables in a given meson system which should be 
contrasted with most NP scenarios outside the MFV framework.  Indeed, if we concentrate 
on $B_s^0-\bar B_s^0$ mixing, $b\to s\mu^+\mu^-$ and $b\to s \nu\bar\nu$ 
observables there are only four new parameters to our disposal: the 
three {\it reduced couplings} of which the first one is generally complex 
and the other two real. These are  (our normalizations of couplings are given 
in Section~\ref{sec:2})
\be\label{Step3a}
\bar\Delta_L^{sb}(Z')=\frac{\Delta_L^{sb}(Z')}{M_{Z'}}, \quad
\bar\Delta_A^{\mu\bar\mu}(Z')=\frac{\Delta_A^{\mu\bar\mu}(Z')}{ M_{Z'}}, 
\quad \bar\Delta_V^{\mu\bar\mu}(Z')=\frac{\Delta_V^{\mu\bar\mu}(Z')}{ M_{Z'}},
\ee
where the {\it bar} distinguishes these couplings from the ones used in 
 \cite{Buras:2012jb}. The couplings $\Delta_{A,V}^{\mu\bar\mu}(Z')$ are defined 
in (\ref{DeltasVA}) and due to $SU(2)_L$ symmetry implying in LHS 
$\Delta_{L}^{\nu\bar\nu}(Z')=\Delta_{L}^{\mu\bar\mu}(Z')$ one also has
\be\label{SU2}
\Delta_{L}^{\nu\bar\nu}(Z')=\frac{\Delta_V^{\mu\bar\mu}(Z')-\Delta_A^{\mu\bar\mu}(Z')}{2}.
\ee
Concrete models satisfying this relation are the 3-3-1 models analyzed in 
\cite{Buras:2012dp}. This relation has also been emphasized recently in 
\cite{Altmannshofer:2013foa,Gauld:2013qba}.

The four new parameters in (\ref{Step3a}) describe in this model NP effects 
in flavour violating processes, in 
particular
\be\label{Class7} 
\Delta M_s, \quad S_{\psi \phi}, \quad B_s\to\mu^+\mu^-, \quad S_{\mu\mu}^s, \quad
B\to K \nu \bar \nu, \quad B\to K^* \nu \bar \nu, \quad B\to X_s \nu \bar \nu.
\ee
and 
\be\label{Class9}
B_d\to K \mu^+\mu^-, \quad B_d\to K^*\mu^+\mu^-,\quad  B_d\to X_s \mu^+\mu^-.
\ee

Extending these considerations to $B_d^0-\bar B_d^0$ mixing and 
$B_d\to\mu^+\mu^-$ we have to our disposal presently 
\be\label{Class6} 
\Delta M_d, \quad S_{\psi K_S}, \quad  B_d\to\mu^+\mu^-.
\ee
It should be noted that in these three observables only $\bar\Delta_L^{db}(Z')$ 
is new as the muon couplings $\bar\Delta_{A,V}^{\mu\bar\mu}(Z')$ are already determined through the observables (\ref{Class7}) and (\ref{Class9}).

In \cite{Buras:2012jb} a very detailed analysis of the correlations among 
observables in (\ref{Class7}) and among the ones in  (\ref{Class6}) has been presented taking into account 
the constraints from the processes (\ref{Class9}) known at the time of
 our analysis. 
In the meantime two advances on the experimental side have been made that deal with processes listed above:
\begin{itemize}
\item
The LHCb and CMS collaborations presented new results on $B_{s,d}\to\mu^+\mu^-$
\cite{Aaij:2013aka,Chatrchyan:2013bka,CMS-PAS-BPH-13-007}.  While the 
branching ratio for $B_s\to\mu^+\mu^-$ turns out to be rather close to SM prediction, although a bit lower, 
the central value for the one of $B_d\to\mu^+\mu^-$ is by a factor of 3.5 higher  than its SM value.
\item
LHCb collaboration reported new results on angular observables in 
$B_d\to K^*\mu^+\mu^-$ that show significant departures from SM expectations 
\cite{Aaij:2013iag,Aaij:2013qta}. Moreover, new data on the observable $F_L$, 
consistent with LHCb value in \cite{Aaij:2013iag} have been presented by 
CMS \cite{Chatrchyan:2013cda}.
\end{itemize}
In particular the anomalies in $B_d\to K^*\mu^+\mu^-$ triggered recently 
 two sophisticated analyses \cite{Descotes-Genon:2013wba,Altmannshofer:2013foa} 
with the goal to understand 
the data and to indicate what type of new physics could be responsible 
for these departures from the SM. Both analyses point 
toward NP contributions in the
modified coefficients  $C_{7\gamma}$ and $C_{9}$ with the following shifts with
respect to their SM values: 
\be
C^{\rm NP}_{7\gamma} < 0, \qquad C^{\rm NP}_{9} < 0.
\ee
Other possibilities, in particular involving right-handed currents, have been discussed in \cite{Altmannshofer:2013foa}.
It should be emphasized at this point that these analyses are subject 
to theoretical uncertainties, which have been discussed at length in 
\cite{Khodjamirian:2010vf,Beylich:2011aq,Matias:2012qz,Jager:2012uw,Descotes-Genon:2013wba} and it remains to be seen whether the observed anomalies are only 
result of statistical fluctuations and/or underestimated error uncertainties. 
Assuming that this is not the case we will investigate how LHS faces these 
data.

As far as $C^{\rm NP}_{9}$ is concerned, the favorite scenario suggested in
\cite{Descotes-Genon:2013wba}  is precisely the LHS scenario analyzed in \cite{Buras:2012jb} but with a simplifying assumption that $ C^{\rm
NP}_{9}$ is real. 
In 
\cite{Descotes-Genon:2013wba} it has also been suggested that 
$\Delta_A^{\mu\bar\mu}(Z')\approx0$ and in fact also in the examples of  $Z^\prime$ models presented in \cite{Altmannshofer:2013foa,Gauld:2013qba} the axial-vector 
coupling has been set to zero. Clearly such a solution, as already mentioned in 
these papers,
would eliminate NP contributions to $B_s\to\mu^+\mu^-$ which although  
consistent with the present data is not particularly interesting. We would like to add that such a choice 
would also eliminate NP contributions to $B_d\to\mu^+\mu^-$ precluding 
the explanation in LHS of a possible enhancement of $\mathcal{B}(B_d\to\mu^+\mu^-)$ indicated by the LHCb and CMS data.

It should be remarked that according to the analysis in 
\cite{Altmannshofer:2013foa} $ C^{\rm NP}_{9}$, while reducing significantly 
the anomalies in the angular observables $S_5$ and $F_L$, cannot provide 
a complete solution when the data on $B\to K\mu^+\mu^-$ and the forward-backward asymmetry $A_{\rm FB}$ are taken into account. 
Yet the nice pattern that 
a negative  $ C^{\rm NP}_{9}$ automatically shifts  $S_5$ and $F_L$ in the 
right direction towards the data is a virtue of this simple scenario.

The inclusion of a negative NP contribution $C^{\rm NP}_{7\gamma}$, which exhibits 
the same pattern in the shifts in  $S_5$ and $F_L$, together with $ C^{\rm NP}_{9}$ would provide 
a better fit to the data. However in the context of our general analysis in 
\cite{Buras:2012jb} we have demonstrated that the contribution of $Z^\prime$ to $C_{7\gamma}$ is fully negligible. Whether this is 
a problem for LHS remains to be seen when the data on $B_d\to K^*\mu^+\mu^-$ and $B\to X_s\gamma$ improve. 
Thus indeed in what follows we can concentrate on modifications in two Wilson coefficients, $C_9$ and $C_{10}$, that are relevant for 
flavour observables in $B_d\to K^*\mu^+\mu^-$ and $B_s\to\mu^+\mu^-$, respectively.

Having developed the full machinery for analyzing the processes in question in 
$Z^\prime$ models in \cite{Buras:2012jb} we would like in this 
paper to have still another look at the LHS scenario in view of 
the  most recent data. As already two detailed analyses of anomalies
in  $B_d\to K^*\mu^+\mu^-$ have been presented
in \cite{Descotes-Genon:2013wba,Altmannshofer:2013foa}, our paper will be 
dominated by $B_{s,d}\to\mu^+\mu^-$ decays. Therefore, in contrast to these 
papers the vector-axial coupling $\Delta_A^{\mu\bar\mu}(Z')$ will play a crucial 
role in our analysis.
In particular in the spirit of  \cite{Buras:2013ooa,Buras:2012jb} we will
expose in the LHS the correlations between $\Delta F=2$ observables and  $B_{s,d}\to\mu^+\mu^-$ illustrating their dependence on $\Delta
M_{s,d}/(\Delta M_{s,d})_{\rm SM}$, $S_{\psi\phi}$ and $S_{\psi K_S}$ which should be precisely determined in  this decade. Here the
theoretically clean CMFV relation 
between these observables \cite{Buras:2003td}, that is violated in the LHS, 
will play a prominent role. However, we will also briefly discuss the correlation 
between $\Delta M_s$ and the size of  $C^{\rm NP}_{9}$ necessary to 
understand the $B_d\to K^*\mu^+\mu^-$ anomalies \cite{Descotes-Genon:2013wba,Altmannshofer:2013foa}.

Now, in  \cite{Buras:2012jb} we have performed already 
a very detailed analysis 
of 
the processes and observables listed in (\ref{Class7}) and  (\ref{Class6}) in the LHS scenario and it is mandatory for us to state what is new in the present paper:
\begin{itemize}
\item
First of all the data on $B_{s,d}\to\mu^-\mu^-$ changed relative to those 
known at the time of the analysis in  \cite{Buras:2012jb} and we would like 
to confront LHS with these data. In particular  as stated above we will calculate the deviations from  the stringent CMFV relation between $\Delta M_{s,d}$ and $\mathcal{B}(B_{s,d}\to\mu^+\mu^-)$ \cite{Buras:2003td} present in this model that has not been done in  \cite{Buras:2012jb} nor in any other paper known to us.
An exception is our analysis of 3-3-1 models in \cite{Buras:2012dp} but in 
this concrete $Z^\prime$ model NP effects in $B_d\to\mu^+\mu^-$ are too small 
to reproduce the recent data  within $1\sigma$.
\item
While in  \cite{Buras:2012jb} we have demonstrated how the coupling $\Delta_L^{sb}(Z')$ could be determined from $\Delta M_s$, $S_{\psi\phi}$ and 
$B_s\to\mu^+\mu^-$ observables and the coupling 
$\Delta_L^{db}(Z')$ from $\Delta M_d$, $S_{\psi K_S}$ and
$B_d\to\mu^+\mu^-$,  we have done it for chosen values of the muon 
coupling $\Delta_A^{\mu\bar\mu}(Z')$ and $M_{Z^\prime}$ and using in particular 
the CP-asymmetry $S_{\mu\mu}^s$ which is very difficult to measure. In the 
present paper we want to summarize how the three reduced couplings listed in 
(\ref{Step3a}) can be determined by invoking in addition the angular observables in 
$B_d\to K^*\mu^+\mu^-$  that can be much easier measured than $S_{\mu\mu}^s$ and 
not making a priori any assumptions on $\Delta_{A,V}^{\mu\bar\mu}(Z')$ and $M_{Z^\prime}$
\item
In view of the continued progress in lattice calculations we will 
investigate how the results presented here depend on the values of
$C_{B_s}$ and $C_{B_d}$ whose departure from unity measures the NP effects 
in $\Delta M_s$ and $\Delta M_d$. We will see that precise knowledge 
of these parameters as well as precise measurements of CP-asymmetries 
$S_{\psi\phi}$ and $S_{\psi K_S}$ are very important for the determination 
of the couplings in (\ref{Step3a}).
\item
We will refine our previous analysis of the correlations of $B_s\to \mu^+\mu^-$ and $b\to s\nu\bar\nu$ observables. 
\item
In the context of our presentation we will critically analyze the LHS 
scenario with a real coefficient $C^{\rm NP}_9$ advocated in \cite{Descotes-Genon:2013wba,Altmannshofer:2013foa}. In particular we present a correlation between real $C^{\rm NP}_9$ and $\Delta M_s$ pointing out that in LHS $\Delta M_s$ is then uniquely enhanced which could be tested one day when lattice calculations 
and values of CKM parameters will be more precise. We discuss briefly the 
implications of such a scenario for $B_{s}\to\mu^+\mu^-$, $S_{\psi\phi}$ 
and $b\to s\nu\bar\nu$ transitions.
\item
  We point out a number of correlations 
between angular observables in $B_d\to K^*\mu^+\mu^-$ which arise in 
LHS, RHS, LRS and ALRS scenarios for couplings of \cite{Buras:2012jb} when 
new CP-violating phases are neglected.
 \item
 As far as $Z$-scenario is concerned we note that large enhancement 
of $\mathcal{B}(B_d\to\mu^+\mu^-)$ found by us in \cite{Buras:2012jb} is fully 
consistent with the recent LHCb and CMS data. However, this scenario does 
not allow the explanation of the $B_d\to K^*\mu^+\mu^-$ anomalies
when the constraint from  $\Delta M_s$ is taken 
into account. Due to the smallness of the vector coupling of $Z$ to muons 
the required modification of $C_9$ implies in this scenario shifts in  
$\Delta M_s$ and in $C_{10}$ that are by far too large.
\end{itemize}

Our paper is organized as follows. In Section~\ref{sec:2} we summarize 
the basic formulae used in our analysis referring often
to the expressions in \cite{Buras:2012jb}, where the same notation is used.
In Section~\ref{sec:3} we show a simple procedure for the determination of 
the reduced couplings in (\ref{Step3a}) up to their signs. In Section~\ref{sec:4},  the most important section of our paper, we perform an
anatomy 
of correlations between $B^0_{s,d}-\bar B^0_{s,d}$ and $B_{s,d}\to\mu^+\mu^-$ 
observables taking into account the information from $B_d\to K^*\mu^+\mu^-$ 
decay.
We also include $b\to s\nu\bar\nu$ in this discussion. 
In Section~\ref{sec:4a} we consider the case of the SM $Z$ gauge boson 
with FCNC couplings. In this case the leptonic reduced couplings are fixed. 
In Section~\ref{sec:5} we address the case of a real $C_9$, 
 the enhancement of $\Delta M_s$ in this case and of the implications for 
other observables. We also discuss briefly scenarios in which $Z^\prime$ and $Z$ have  also right-handed FCNC couplings and point out a
number of correlations 
between angular observables in $B_d\to K^*\mu^+\mu^-$ advertised above.
We conclude in Section~\ref{sec:6}.

Before starting our presentation let us realize that the challenges the LHS 
scenario considered in our paper has to face 
are non-trivial due to the following facts.

The important actors in our paper are the couplings
\be\label{actors}
\bar\Delta_L^{sb}(Z'),  \qquad \bar\Delta_L^{db}(Z'), \qquad
\bar\Delta_A^{\mu\bar\mu}(Z'), \qquad
\bar\Delta_V^{\mu\bar\mu}(Z'),
\ee
in terms of which the decays and related observables in 
(\ref{Class7})-(\ref{Class6}) should be simultaneously described. In view 
of the pattern of the present data mentioned above this is certainly non-trivial for the 
following reasons:
\begin{itemize}
\item
$\bar\Delta_L^{sb}(Z')$  enters both  $B_s\to \mu^+\mu^-$ and  $B_d\to K^*\mu^+\mu^-$ in which NP effects have been found to be small and sizable, respectively. 
This implies through the relation (\ref{AV}) that 
$\bar\Delta_A^{\mu\bar\mu}(Z')<\bar\Delta_V^{\mu\bar\mu}(Z')$.
\item 
The smallness of $\bar\Delta_L^{sb}(Z')$ is welcome as then also NP effects 
in $\Delta M_s$ are small as seen in the data. But then $\bar\Delta_V^{\mu\bar\mu}(Z')$ must be sufficiently large in order to describe the anomalies in
$B_d\to K^*\mu^+\mu^-$.
\item
Similarly  $\bar\Delta_A^{\mu\bar\mu}(Z')$ cannot be  small,  in spite of 
  $B_s\to \mu^+\mu^-$ being SM-like
as otherwise the enhancement of $B_d\to\mu^+\mu^-$ branching ratio over 
SM expectation indicated by the LHCb and CMS data cannot be accommodated. 
Here the sizable coupling $\bar\Delta_L^{db}(Z')$ could help, but it is 
constrained by $\Delta M_d$ and $S_{\psi K_S}$. 
\item
 In the $Z$-scenario the challenges are even larger as the lepton couplings 
are fixed. 
\end{itemize}

We are now ready to investigate how LHS $Z^\prime$ and $Z$ scenarios face
these challenges.

\section{Basic formulae}\label{sec:2}
\subsection{Basic Lagrangian}
The basic formalism for our analysis has been developed in \cite{Buras:2012jb} 
and we collect here only those formulae of that paper that are essential 
for our presentation expressing them this time in terms of the reduced 
couplings in (\ref{Step3a}). However we recall first the basic Lagrangian 
in terms of the couplings used in \cite{Buras:2012jb} ($q=d,s$): 
\be
{\mathcal L}_{\rm FCNC}^{\rm quarks}(Z^\prime)=\left[\bar q\gamma_\mu P_L b \Delta_L^{qb}(Z^\prime)
+\bar q\gamma_\mu P_R b \Delta_R^{qb}(Z^\prime)+h.c.\right] Z^{\prime \mu},
\ee

\be
{\mathcal L}^{\rm leptons}(Z^\prime)=\left[\bar\mu\gamma_\mu P_L \mu \Delta_L^{\mu\bar\mu}(Z^\prime)
+\bar\mu\gamma_\mu P_R \mu\Delta_R^{\mu\bar\mu}(Z^\prime)+ 
\bar\nu\gamma_\mu P_L \Delta_L^{\nu\bar\nu}(Z^\prime)\right] Z^{\prime \mu}
\ee
where $R$ and $L$ stand for right-handed and left-handed couplings $\gamma_\mu(1\pm\gamma_5)/2$. 
Moreover
\be
\bar\Delta_L^{bq}(Z^\prime)=\left[\bar\Delta_L^{qb}(Z^\prime)\right]^*
\ee
and the vector and axial-vector couplings to muons are given as follows
\begin{align}\label{DeltasVA}
\begin{split}
 &\Delta_V^{\mu\bar\mu}(Z')= \Delta_R^{\mu\bar\mu}(Z')+\Delta_L^{\mu\bar\mu}(Z'),\\
&\Delta_A^{\mu\bar\mu}(Z')= \Delta_R^{\mu\bar\mu}(Z')-\Delta_L^{\mu\bar\mu}(Z').
\end{split}
\end{align}
The relation of these couplings to the ones used in \cite{Altmannshofer:2013foa} is as follows
\be
\bar\Delta_{L,R}^{qb}(Z^\prime)=\frac{g_2}{2\cos\theta_W}g_{qb}^{L,R},\qquad 
\Delta_{V,A}^{\mu\bar\mu}(Z')=\frac{g_2}{\cos\theta_W}g_{\mu}^{V,A},
\ee
where $g_2$ is the $SU(2)_L$ gauge coupling. For completeness and because 
of a brief discussion in Section~\ref{sec:5} we have included here right-handed 
couplings $\bar\Delta_{R}^{qb}(Z^\prime)$ which vanish in the LHS.

On the other hand  we do not make any assumptions about diagonal couplings of 
$Z^\prime$ to quarks but we expect them to be non-vanishing. The flavour 
violating couplings in the quark mass eigenstate basis can
e.g.  arise from the non-universality of the diagonal couplings 
in the flavour basis but  other dynamical mechanisms for the FCNC 
couplings in question are possible \cite{Langacker:2008yv}.  Without 
a concrete model it is not possible to establish a relation between 
diagonal and non-diagonal couplings. For a recent discussion see \cite{Guadagnoli:2013mru} and references therein.

\boldmath
\subsection{$\Delta F=2$ observables}
\unboldmath
The $B_s^0-\bar B_s^0$ observables are fully described in LHS by the function 
\begin{equation}\label{Seff}
S(B_s)=S_0(x_t)+\Delta S(B_s)\equiv|S(B_s)|e^{-i2\varphi_{B_s}},
\end{equation}
where $S_0(x_t)$ is the real one-loop SM box function and the additional generally complex term, denoted in 
\cite{Buras:2012jb} by $[\Delta S(B_s)]_{\rm VLL}$,  is the tree-level $Z^\prime$ contribution 
\be\label{Zprime1}
\Delta S(B_s)=
\left[\frac{\bar\Delta_L^{bs}(Z^\prime)}{V^*_{tb}V_{ts}}\right]^2
\frac{4\tilde r}{g_{\text{SM}}^2},\qquad g_{\text{SM}}^2=4\frac{G_F}{\sqrt 2}\frac{\alpha}{2\pi\sin^2\theta_W}.
\ee
Here $\tilde r$ is a QCD factor that includes QCD renormalization group effects 
between $\mu=M_{Z^\prime}$ and $\mu=m_t$ and the difference in matching conditions between full and effective theories in the tree-level
$Z^\prime$ exchanges \cite{Buras:2012fs} and SM box diagrams \cite{Buras:1990fn}. Explicit expression for  $\tilde r$ has been given in
\cite{Buras:2012dp}. One finds $\tilde r=0.985$, $\tilde r=0.953$ and $\tilde r = 0.925$ for $M_{Z^\prime} =1,~3, ~10\tev$, respectively.

The two observables of interest, $\Delta M_s$ and $S_{\psi\phi}$ are then 
given by
\be\label{DMs}
\Delta M_s =\frac{G_F^2}{6 \pi^2}M_W^2 m_{B_s}|V^*_{tb}V_{ts}|^2   F_{B_s}^2\hat B_{B_s} \eta_B |S(B_s)|\,
\ee
and
\begin{equation}
S_{\psi\phi} =  \sin(2|\beta_s|-2\varphi_{B_s})\,, \qquad  V_{ts}=-\vts e^{-i\beta_s}.
\label{eq:3.44}
\end{equation}
with $\beta_s\simeq -1^\circ\,$.

In the case of $B_d^0$ system the corresponding formulae are obtained from 
(\ref{Seff})-(\ref{DMs}) by replacing $s$ by $d$. Moreover (\ref{eq:3.44}) 
is replaced by
\begin{equation}
S_{\psi K_S} =  \sin(2\beta-2\varphi_{B_d})\,, \qquad  V_{td}=\vtd e^{-i\beta}.
\label{eq:3.45}
\end{equation}
The value of $\beta$ depends strongly on $\vub$  but only weakly on its 
phase $\gamma$. For $\gamma=68^\circ$ we find $\beta=21.2^\circ$ and $\beta=25.2^\circ$ for
$\vub=3.4\times 10^{-3}$ and $\vub=4.0\times 10^{-3}$, respectively.

It should be noted that $M_{Z^\prime}$ is hidden in the reduced $Z^\prime bs$ 
coupling and appears explicitly only in $\tilde r$ but this dependence is 
only logarithmic and can be neglected in view of present theoretical and 
experimental uncertainties but should be taken into account in the 
flavour precision era. However, except for this weak dependence 
at tree level it is not possible to measure $M_{Z^\prime}$ through FCNC processes unless the relevant couplings are predicted in a given
model. On the other hand it could be in principle possible through loop processes one day or through direct high energy experiments that
would discover $Z^\prime$.
\boldmath
\subsection{$b\to s \mu^+\mu^-$ observables}
\unboldmath
The two Wilson coefficients that receive NP contributions in LHS model are 
$C_9$ and $C_{10}$. We decompose them into the SM and NP contributions\footnote{These coefficients are defined as in \cite{Buras:2012jb} and the same 
definitions are used in  \cite{Descotes-Genon:2013wba,Altmannshofer:2013foa}.}:
\be
C_9=C_9^{\rm SM}+C_9^{\rm NP},\qquad C_{10}=C_{10}^{\rm SM}+C_{10}^{\rm NP}.
\ee
Then  \cite{Buras:2012jb}\footnote{The quantities $\eta_Y$, $\eta_B$ and $\eta_X$ appearing in the text are QCD corrections for which the values, all $\ord(1)$, can be found in  \cite{Buras:2012jb}.}
\begin{align}
 \sin^2\theta_W C^{\rm SM}_9 &=\sin^2\theta_W P_0^{\rm NDR}+ [\eta_Y Y_0(x_t)-4\sin^2\theta_W Z_0(x_t)],\\
   \sin^2\theta_W C^{\rm SM}_{10} &= -\eta_Y Y_0(x_t)
 \end{align}
 so that 
\be
C^{\rm SM}_9\approx 4.1, \qquad C^{\rm SM}_{10}\approx -4.1~.
\ee
NP contributions have a very simple structure
\begin{align}
 \sin^2\theta_W C^{\rm NP}_9 &=-\frac{1}{g_{\text{SM}}^2}
\frac{\bar\Delta_L^{sb}(Z')\bar\Delta_V^{\mu\bar\mu}(Z')} {V_{ts}^* V_{tb}} ,\label{C9}\\
   \sin^2\theta_W C^{\rm NP}_{10} &= -\frac{1}{g_{\text{SM}}^2}
\frac{\bar\Delta_L^{sb}(Z')\bar\Delta_A^{\mu\bar\mu}(Z')}{V_{ts}^* V_{tb}}\label{C10}.
 \end{align}
 and consequently we have an important relation
\be\label{AV}
\frac{ C^{\rm NP}_{10}}{ C^{\rm NP}_{9}}=\frac{\bar\Delta_A^{\mu\bar\mu}(Z')}{\bar\Delta_V^{\mu\bar\mu}(Z')},
\ee
which involves only leptonic couplings.

$Y_0(x_t)$ and $Z_0(x_t)$ are   SM one-loop functions, analogous to $S_0(x_t)$. Explicit expressions for them can be found in \cite{Buras:2012jb}. 
$C_{10}$ is scale independent as far as pure QCD corrections are 
concerned but at higher order in QED the relevant operator mixes with other 
operators \cite{Bobeth:2003at,Huber:2005ig}. This effect will be included 
in the complete calculation of NLO electroweak corrections to $B_s\to\mu^+\mu^-$. $C_9$ is affected by QCD corrections, 
present in the term   $P_0^{\rm NDR}$,
through mixing with four-quark current-current operators. Its value is usually 
quoted at $\mu=\ord(m_b)$. Beyond one-loop this term is  renormalization scheme dependent but as demonstrated in \cite{Buras:1994dj} at the NLO 
level this dependence is canceled by QCD corrections to the matrix 
elements of the relevant operators. By now these corrections are known at the 
NNLO level \cite{Bobeth:1999mk,Bobeth:1999ww,Bobeth:2003at} and are taken into account in the extraction of $C_9^{\rm NP}$ from the data.
Finally, it should be mentioned that there are also QCD corrections affecting 
the NP part due to the mixing of new four-quark operators generated through $Z^\prime$ exchange. These corrections would effectively modify the term $P_0^{\rm NDR}$. 
Corrections of this type have been calculated in the case of $Z^\prime$ 
contributions to $B\to X_s\gamma$ decay in \cite{Buras:2011zb,Buras:2012jb} and found to 
be small. As the anomalous dimensions in the present case are smaller than 
in the case of $B\to X_s\gamma$ it is safe to neglect these corrections.

One has then in the case of $B_{s}\to\mu^+\mu^-$ decay  \cite{deBruyn:2012wk,Fleischer:2012fy,Buras:2013uqa}
\be\label{Rdef}
\frac{\overline{\mathcal{B}}(B_{s}\to\mu^+\mu^-)}{\overline{\mathcal{B}}(B_{s}\to\mu^+\mu^-)_{\rm SM}}
= \left[\frac{1+{\cal A}^{\mu\mu}_{\Delta\Gamma}\,y_s}{1+y_s} \right] |P|^2, 
\qquad
P=\frac{C_{10}}{C_{10}^{\rm SM}}
\equiv |P|e^{i\varphi_P},
\ee
where
\be 
{\cal A}^{\mu\mu}_{\Delta\Gamma}=\cos(2\varphi_P-2\varphi_{B_s}),\qquad
	y_s\equiv\tau_{B_s}\frac{\Delta\Gamma_s}{2}
=0.088\pm0.014.
\ee
The {\it bar} indicates that $\Delta\Gamma_s$ effects have been taken into account. In the SM and CMFV ${\cal A}^{\mu\mu}_{\Delta\Gamma}=1$ 
but in the LHS it is 
slightly smaller and we take this effect into account. Generally as shown 
in \cite{Buras:2013uqa} ${\cal A}^{\mu\mu}_{\Delta\Gamma}$ can serve to test 
NP models as it can be determined in time-dependent 
measurements \cite{deBruyn:2012wk,Fleischer:2012fy}.
Of interest is also the CP asymmetry
\begin{equation}\label{defys}
S^s_{\mu\mu}=\sin(2\varphi_P-2\varphi_{B_s}),
\end{equation}
which has been studied in detail in \cite{Buras:2012jb,Buras:2013uqa} 
in the context of $Z^\prime$ models.

 In the case of $B_d\to\mu^+\mu^-$ decay the formulae given above apply with 
$s$ replaced by $d$ and $y_d\approx 0$. While $C_{10}^{\rm SM}$ remains unchanged, 
$C_{10}^{\rm NP}$ is clearly modified through the replacement of $V_{ts}$ by 
$V_{td}$ and different $Z^\prime$ coupling to quarks. But the muon coupling 
remains unchanged and this will allow a correlation between $B_s\to\mu^+\mu^-$ 
and  $B_d\to\mu^+\mu^-$ which will be investigated within LHS here for 
the first time. Explicit formulae for $B_d\to\mu^+\mu^-$ can be found in 
\cite{Buras:2012jb}.

Concerning the  status 
 of the branching ratios for 
$B_{s,d}\to \mu^+\mu^-$ decays we have  
\be\label{LHCb2}
\overline{\mathcal{B}}(B_{s}\to\mu^+\mu^-)_{\rm SM}= (3.56\pm0.18)\cdot 10^{-9},
\quad
\overline{\mathcal{B}}(B_{s}\to\mu^+\mu^-) = (2.9\pm0.7) \times 10^{-9}, 
\ee
\be\label{LHCb3}
\mathcal{B}(B_{d}\to\mu^+\mu^-)_{\rm SM}=(1.05\pm0.07)\times 10^{-10}, \quad
\mathcal{B}(B_{d}\to\mu^+\mu^-) =\left(3.6^{+1.6}_{-1.4}\right)\times 10^{-10}, \quad
\ee
where the SM values are based on  \cite{Buras:2012ru,Buras:2013uqa} and experimental data are the most recent average of the results from LHCb and CMS
\cite{Aaij:2013aka,Chatrchyan:2013bka,CMS-PAS-BPH-13-007}.

In the case of $B\to K^*\mu^+\mu^-$ we will concentrate our discussion on 
the Wilson coefficient $C_9^{\rm NP}$ which can be extracted from 
the angular observables, in particular $\langle F_L\rangle$, $\langle S_5\rangle$ and $\langle A_8 \rangle$, 
in which within the LHS  NP contributions enter 
exclusively through this coefficient. On the other hand $\IM(C_{10}^{\rm NP})$ 
governs the CP-asymmetry $\langle A_7 \rangle$. Useful approximate 
expressions for these four angular observables in terms of  $C_9^{\rm NP}$  
and  $C_{10}^{\rm NP}$  have been provided in \cite{Altmannshofer:2013foa}.

The recent $B\to K^*\mu^+\mu^-$ anomalies imply the following 
ranges for  $C_9^{\rm NP}(B_s)$  \cite{Descotes-Genon:2013wba,Altmannshofer:2013foa} respectively
\be\label{ANOM}
 C_9^{\rm NP}(B_s)=-(1.6\pm0.3), \qquad 
 C_9^{\rm NP}(B_s)=-(0.8\pm0.3)  
\ee
As 
 $C_9^{\rm SM}(B_s)\approx 4.1$ at $\mu_b=4.8\gev$, these are very significant 
suppressions of this coefficient.  We note that $C_9$ remains real as in the 
SM.  We will have a closer look at the implications of this result for 
both values quoted above. The details behind these two results that 
differ by a factor of two is discussed in \cite{Altmannshofer:2013foa}. 
In fact inspecting Figs.~3 and 4 of the latter paper one sees that if 
the constraints from $A_{\rm FB}$ and $B\to K\mu^+\mu^-$ were not taken into account $C_9^{\rm NP}(B_s)\approx -1.4$ could alone explain the anomalies in 
the observables $F_L$ and $S_5$. But the inclusion of these constraints 
 reduces the size of this coefficient. Yet values of  
$C_9^{\rm NP}(B_s)\approx -(1.2-1.0)$ seem to give reasonable agreement with 
all data and the slight reduction of departure of $F_L$ and $S_5$ from 
their SM values in the future data would allow to explain the two anomalies 
with the help of  $C_9^{\rm NP}(B_s)$ as suggested originally in 
 \cite{Descotes-Genon:2013wba}.

Further support for this picture comes from our analysis below and
a very recent comprehensive Bayesian analysis of
the authors of \cite{Beaujean:2012uj,Bobeth:2012vn} in \cite{Beaujean:2013soa},
that appeared after our paper.  They find that although
SM works well, if one wants to interpret the data in extensions of the SM
then NP scenarios with dominant NP effect in $C_9$ are favoured although
the inclusion of chirality-flipped operators in agreement with \cite{Altmannshofer:2013foa} would help to reproduce the data. This is also
confirmed in
in the very recent paper in
in \cite{Horgan:2013pva}.
References to earlier papers on  $B\to K^*\mu^+\mu^-$
 by all these authors can be found in  \cite{Descotes-Genon:2013wba,Altmannshofer:2013foa,Bobeth:2012vn} and \cite{Buras:2013ooa}.

\boldmath
\subsection{Correlations between $\Delta M_{s,d}$, $B_{s,d}\to\mu^+\mu^-$ and $C_9^{\rm NP}$}
\unboldmath
In \cite{Buras:2012jb} a number of correlations between $\Delta F=2$ and $\Delta F=1$ observables in LHS have been identified. Here we want to concentrate on the correlations  between $\Delta M_{s,d}$, $B_{s,d}\to\mu^+\mu^-$ and $C_9^{\rm NP}$
as they can be exposed analytically.

First the CMFV relation between  
$\mathcal{B}(B_q\to\mu^+\mu^-)$ and   $\Delta M_{s,d}$ \cite{Buras:2003td} 
generalizes in LHS to $(q=s,d)$
\begin{align}\label{NonDirect}
& \mathcal{B}(B_q\to\mu^+\mu^-) = C \frac{\tau_{B_q}}{\hat B_q}\frac{|Y_A^q|^2}{|S(B_q)|}\Delta M_q,\\
&\text{with}\quad C = 6\pi \frac{\eta_Y^2}{\eta_B^2}\left(\frac{\alpha}{4\pi \sin^2\theta_W}\right)^2\frac{m_\mu^2}{M_W^2}= 4.395\cdot 10^{-10},
\end{align}
$S(B_q)$ given in (\ref{Seff}) and 
\be 
Y_A^q= \eta_Y Y_0(x_t)
+\frac{\Delta_L^{qb}(Z^\prime)}{V_{tb}V^*_{tq}}\frac{\left[\Delta_A^{\mu\bar\mu}(Z^\prime)\right]}{M_{Z^\prime}^2g_\text{SM}^2}.
\ee

Note that these relations are free 
from $F_{B_q}$  dependence  but in contrast to CMFV they depend on 
 $V_{tq}$  as generally $\Delta_L^{qb}(Z^\prime)$ are not aligned with 
$V_{tb}V^*_{tq}$. The main uncertainty in these relations 
comes from the parameters 
 ${\hat B_q}$ that are known presently with an accuracy of $\pm 8\%$ and $\pm 4.5\%$ for $\hat B_d$ and $\hat B_s$, respectively (see new version of FLAG 
\cite{Colangelo:2010et}).
 More 
accurate is the relation \cite{Buras:2003td} 
 \be\label{CMFV6}
 \frac{\mathcal{B}(B_{s}\to\mu^+\mu^-)}{\mathcal{B}(B_{d}\to\mu^+\mu^-)}
 =\frac{\hat B_{d}}{\hat B_{s}}
 \frac{\tau( B_{s})}{\tau( B_{d})} 
 \frac{\Delta M_{s}}{\Delta M_{d}}r, \qquad
 r=\left|\frac{Y_A^s}{Y_A^d}\right|^2\left|\frac{S(B_d)}{S(B_s)}\right|, \qquad 
\frac{\hat B_{d}}{\hat B_{s}}=0.99\pm0.02
\ee
where the departure of $r$ from unity measures effects which go beyond CMFV. 
Still as shown already in  \cite{Buras:2003td} in the context of supersymmetric  models with MFV  and in the context of 
GMFV in \cite{Kagan:2009bn} at large $\tan\beta$ one also finds 
$r\approx 1$ so that 
this relation offers also a test of these scenarios. On the other hand 
the most general test of MFV, as emphasized in \cite{Hurth:2008jc}, is the 
proportionality of the ratio of the two branching ratios in question to $\vts^2/\vtd^2$.

It should be noted that in (\ref{CMFV6})
the only theoretical uncertainty enters through
the ratio $\hat B_{s}/\hat B_{d}$
 that is already now known from  lattice calculations with impressive 
accuracy of roughly $\pm 2\%$ \cite{Carrasco:2013zta} as given in (\ref{CMFV6})\footnote{This result is not included in the recent FLAG
update which quotes
$0.95\pm0.10$.}. 
 Therefore the relation (\ref{CMFV6}) should allow
 a precision test of CMFV, related scenarios mentioned above and LHS even if the branching ratios $\mathcal{B}(B_{s,d}\to\mu^+\mu^-)$ would turn out to deviate from SM 
predictions only  by $10-15\%$. We should emphasize that such a precision test is 
not possible with any angular observable in $B_d\to K^*\mu^+\mu^-$ due to 
form factor uncertainties. On the other hand these observables provide more 
information on NP than the relation (\ref{CMFV6}).

In fact as seen in 
Fig.~\ref{fig:BdvsBsLHS} the present data for $r$ differ from its CMFV value ($r=1$) by more than a factor of four 
\be\label{rexp}
r_{\rm exp}=0.23\pm 0.11.
\ee
Even if  in view of large experimental uncertainties one cannot claim that here NP is at work, this plot invites us to investigate whether LHS 
could cope with the future more precise experimental results in which the central values of the branching ratios  in (\ref{LHCb2}) and (\ref{LHCb3})
would not change by much.

As far as  
the Wilson coefficients $C_9^{\rm NP}$ and 
$C_{10}^{\rm NP}$ are concerned we have two important relations
\be\label{SC9}
(\Delta S)^\ast=4\tilde rg^2_{\rm SM}\sin^4\theta_W \left[\frac{C_9^{\rm NP}}{\bar\Delta_V^{\mu\bar\mu}(Z')}\right]^2=0.037
\left[\frac{C_9^{\rm
NP}}{\Delta_V^{\mu\bar\mu}(Z')}\right]^2 \left[\frac{M_{Z^\prime}}{1\tev}\right]^2,
\ee

\be\label{SC10}
(\Delta S)^\ast=4\tilde rg^2_{\rm SM}\sin^4\theta_W \left[\frac{C_{10}^{\rm NP}}{\bar\Delta_A^{\mu\bar\mu}(Z')}\right]^2=0.037
\left[\frac{C_{10}^{\rm NP}}{\Delta_A^{\mu\bar\mu}(Z')}\right]^2 \left[\frac{M_{Z^\prime}}{1\tev}\right]^2,
\ee
which we have written in a form suitable for the analysis in Section~\ref{sec:5}. We recall that $S_{\rm SM}=S_0(x_t)=2.31$.

These relations can be derived from (\ref{Zprime1}), (\ref{C9}) and (\ref{C10}). The last 
relation, already encoded in previous relations, has been extensively studied in \cite{Buras:2012jb} for fixed value 
of the coupling $\bar\Delta_A^{\mu\bar\mu}(Z')$. It is evident that independently  of the sign of this coupling in the case of a real $C_{10}^{\rm NP}$, 
$\Delta S$ and $\Delta M_s$ will be enhanced which with the lattice value $\sqrt{\hat B_{B_s}}F_{B_s}=(279\pm13)\mev$ \cite{Laiho:2009eu} used in \cite{Buras:2012jb} would be a problem \cite{Buras:2013raa}.
Making 
$C_{10}^{\rm NP}$ complex allowed through destructive interference with the 
SM contribution to lower the value of $\Delta M_s$ and bring it to agree with 
the data. With the new value for $\sqrt{\hat B_{B_s}}F_{B_s}\mev$ given below this problem is softened.

Concerning (\ref{SC9}), we note that in the case of a real $C_{10}^{\rm NP}$, the relation (\ref{AV}) implies that also $C_{9}^{\rm NP}$ must be real 
in LHS so that also this relation implies in this case uniquely an enhancement of $\Delta S$ and $\Delta M_s$. 

In the next section we will  get an idea on the range of the values of
 $\bar\Delta_A^{\mu\bar\mu}(Z')$ by looking simultaneously at $B_s\to\mu^+\mu^-$ 
and $B_d\to\mu^+\mu^-$ decays. In the case of $\bar\Delta_V^{\mu\bar\mu}(Z')$ 
the simultaneous consideration of the decays $B_d\to K^*\mu^+\mu^-$ and 
$B_d\to \rho\mu^+\mu^-$ could give us in principle information about 
the size of this coupling. However, there is not enough experimental 
information on the latter decay and such an exercise cannot be performed 
at present.

On the other hand in the context of (\ref{SC9}) it has been noted in \cite{Gauld:2013qba} that 
$\bar\Delta_V^{\mu\bar\mu}(Z')$ could be eliminated in favour of the violation 
of the CKM unitarity in $Z^\prime$ models studied by Marciano and Sirlin long 
time ago \cite{Marciano:1987ja} if one assumes that $\Delta_A^{sb}=0$ 
and the diagonal couplings of $Z^\prime$ to quarks vanish. As already admitted 
by the authors of \cite{Gauld:2013qba} such a model is not realistic.  
Therefore we provide here a more general formula which uses the results in 
\cite{Marciano:1987ja} without making the assumptions made in \cite{Gauld:2013qba}.

We denote the violation of CKM unitarity by
\be
\tilde\Delta_{\rm CKM}=1-\sum_{q=d,s,b}|V_{uq}|^2=-\Delta_{\rm CKM},
\ee
with $\Delta_{\rm CKM}$ used in \cite{Marciano:1987ja,Gauld:2013qba}.
Then  for $M_{Z'}\gg M_W$ we find 
\be\label{Marciano}
\tilde\Delta_{\rm CKM}=\frac{3\sqrt{2}}{8 G_F}\frac{\alpha}{\pi\sin^2\theta_W}
\bar\Delta_L^{\mu\bar\mu}(Z')(\bar\Delta_L^{\mu\bar\mu}(Z')-\bar\Delta_L^{d\bar d}(Z'))\ln\frac{M_{Z'}^2}{M_W^2},
 \ee
where $\bar\Delta_L^{d\bar d}$ is the diagonal coupling of $Z^\prime$ to down-quarks which is assumed to be generation independent.

The triple correlation between $\Delta M_s$, $\tilde\Delta_{\rm CKM}$ and 
$C_9^{\rm NP}$ found in \cite{Gauld:2013qba} only follows for the case
\be\label{asummption}
{\bar\Delta_V^{\mu\bar\mu}(Z')}=\frac{\bar\Delta_L^{\mu\bar\mu}(Z')}{2}, \qquad 
\bar\Delta_L^{d\bar d}(Z')=0,
\ee 
where the first equality is equivalent to $\bar\Delta_A^{\mu\bar\mu}(Z')=0$.
The triple correlation in question can now be rewritten as
\be\label{Triple}
(\Delta S)^\ast\, \tilde\Delta_{\rm CKM}=\frac{3}{4}\tilde r \left(\frac{\alpha}{\pi}\right)^2 (C_9^{\rm NP})^2 \ln\frac{M_{Z'}^2}{M_W^2},
\ee
which allows better to follow the signs than the expression given by these authors. As now $\tilde\Delta_{\rm CKM}\ge0$, it is evident also from this formula that in the case of a real $C_9^{\rm NP}$, independently of its sign,  $\Delta S$ is real and strictly positive enhancing uniquely $\Delta M_s$. In \cite{Gauld:2013qba} only $|\Delta S|$ has been studied.

However, generally the assumptions in (\ref{asummption}) are violated and  
in examples shown in \cite{Marciano:1987ja} there is always a 
quark contribution to $\tilde\Delta_{\rm CKM}$. In fact these authors 
present GUT examples where the quark contribution cancels the one of leptons 
so that in such a model there is no violation of CKM unitarity. 

Independently of this discussion the case of a real  $C_9^{\rm NP}$ is 
interesting in itself and in Section~\ref{sec:5} we will investigate how the predictions of the LHS model would look like 
in the presence of a real $C_9^{\rm NP}$ as large as
required to remove the anomalies 
in the data on $B_d\to K^*\mu^+\mu^-$.

\section{Determining the parameters in the LHS}\label{sec:3}
In principle there are many ways to bound or even determine the reduced 
parameters in (\ref{Step3a}). Here we present one route which simultaneously 
allows already at the early stage to test the LHS. This route could be improved and  modified dependently on evolution of experimental data.

To this end we use the parametrization
\be\label{S23}
\bar\Delta_L^{sb}(Z')=-|\bar\Delta_L^{sb}(Z')|e^{i\delta_{23}}, \qquad        |\bar\Delta_L^{sb}(Z')|=\frac{\tilde s_{23}}{M_{Z'}},
\ee
where $\tilde s_{23}$ and $\delta_{23}$ are parameters used in \cite{Buras:2012jb} and
the minus sign is introduced
to cancel the minus sign in $V_{ts}$ in the relevant phenomenological formulae.
 For $B_d-\bar B_d^0$ mixing and $B_d\to\mu^+\mu^-$ $s$ is replaced by $d$ and 
no minus sign is introduced. Moreover $\delta_{23}$ is replaced by $\delta_{13}$.

\subsubsection*{Step 1}

Measurements of $\Delta M_s$, $S_{\psi\phi}$ and $\Delta\Gamma_s$ determine uniquely $|\bar\Delta_L^{sb}(Z')|$ and two values of the  phase $\delta_{23}$, 
differing by $180^\circ$ corresponding to two oases determined in \cite{Buras:2012jb}: {\it blue} and {\it purple} oasis for low and high $\delta_{23}$, respectively.  Equivalently the phase  $\delta_{23}$ could be fixed through the blue 
oasis. The purple oasis is then reached by just flipping the sign of 
$\bar\Delta_L^{sb}(Z')$. 

The same procedure is applied to the $B_d$-system which determines $\bar\Delta_L^{db}(Z')$. The two values for  the  phase $\delta_{13}$
correspond to 
{\it yellow} and {\it green} oasis in \cite{Buras:2012jb}, respectively. It should be 
recalled that the outcome of this determination depends on the value of $\vub$ 
resulting in two LHS scenarios, LHS1 and LHS2, corresponding to exclusive 
and inclusive determinations of $\vub$. This point is elaborated at length in 
\cite{Buras:2013ooa,Buras:2012jb}.

\subsubsection*{Step 2}

With the result of Step 1 we can immediately perform an important test of the
LHS as in this model in the $b\to s\mu^+\mu^-$ case
\be\label{IMRE}
\frac{\IM(C_9^{\rm NP})}{\RE(C_9^{\rm NP})}=\frac{\IM(C_{10}^{\rm NP})}{\RE(C_{10}^{\rm NP})}=\tan(\delta_{23}-\beta_s).
\ee
Two important points should be noticed here. These two ratios have to be equal to each other. Moreover they do not depend on whether we are in the blue or purple oasis. This 
test is interesting in the context of recent proposals that $C_9^{\rm NP}$ 
could be real. We point out that in the context of LHS this proposal 
can be tested through the correlation 
with $B_s^0-\bar B_s^0$ mixing. We will elaborate on it in Section~\ref{sec:5}.

\subsubsection*{Step 3}

In order to determine $|\bar\Delta_A^{\mu\bar\mu}(Z')|$ we use
$\overline{\mathcal{B}}(B_{s}\to\mu^+\mu^-)$ together with results of Step 1.
The sign of this coupling cannot be determined uniquely  as only the 
product of the couplings $\bar\Delta_L^{sb}(Z')$ and $\bar\Delta_A^{\mu\bar\mu}(Z')$ is measured by $\overline{\mathcal{B}}(B_{s}\to\mu^+\mu^-)$. 
As the sign of $\bar\Delta_L^{sb}(Z')$ could not be fixed in Step 1 also 
the sign of  $\bar\Delta_A^{\mu\bar\mu}(Z')$ cannot be determined. Yet
this ambiguity has no impact on physical implications as the measurements of 
$\Delta M_s$, $S_{\psi\phi}$ and $\overline{\mathcal{B}}(B_{s}\to\mu^+\mu^-)$
correlate the signs of these two couplings so that once the three observables 
have been measured it is irrelevant whether the subsequent calculations 
are performed in the blue or purple oasis. Moving from one oasis 
to the other will require simultaneous  change of the sign of  $\bar\Delta_A^{\mu\bar\mu}(Z')$ to agree with the data on $\overline{\mathcal{B}}(B_{s}\to\mu^+\mu^-)$.
The subsequent predictions 
for $S^s_{\mu\mu}$ and $\langle A_7 \rangle$ and the correlations of these two
observables with $\Delta M_s$, $\overline{\mathcal{B}}(B_{s}\to\mu^+\mu^-)$ 
and $S_{\psi\phi}$ will not be modified in the new oasis.

However when the result for $|\bar\Delta_A^{\mu\bar\mu}(Z')|$ determined 
in this manner is used in conjunction
      with $\bar\Delta_L^{db}(Z')$ obtained in Step 1 to predict 
$\mathcal{B}(B_d\to\mu^+\mu^-)$ the final result will depend on the sign of 
these two couplings. But this difference will be distinguished simply through 
the enhancement or suppression of this branching ratio relative to its 
SM value. Moreover the value of $\mathcal{B}(B_d\to\mu^+\mu^-)$ will depend 
on whether LHS1 or LHS2 is considered.

\subsubsection*{Step 4}

In order to determine $|\bar\Delta_V^{\mu\bar\mu}(Z')|$ we have to know 
$C_9^{\rm NP}$ which can be extracted from one of the following 
angular observables: $\langle F_L\rangle$, $\langle S_5\rangle$ and $\langle A_8 \rangle$. Again only the product  of
$\bar\Delta_L^{sb}(Z')$
and $\bar\Delta_V^{\mu\bar\mu}(Z')$ is determined in this manner 
and consequently the sign of $\bar\Delta_V^{\mu\bar\mu}(Z')$ cannot be 
determined uniquely as was the case of  $\bar\Delta_A^{\mu\bar\mu}(Z')$ in Step 3. Yet, one can convince oneself that the
correlations between these three observables 
and the ones in the $B_s$ meson system considered in previous steps,  do not depend on this sign once the products of couplings in questions have been 
determined uniquely in Step 3 and this step.

\subsubsection*{Step 5}

As the determination of quark and muon couplings is completed we can 
first determine $\bar\Delta_L^{\nu\bar\nu}(Z')$ by means of (\ref{SU2}) 
and subsequently make unique predictions for all $b\to s \nu\bar\nu$ 
transitions. Indeed in this case the relevant one-loop function is 
\be\label{XLB}
 X_{\rm L}(B_s)=\eta_X X_0(x_t)+\left[\frac{\bar\Delta_{L}^{\nu\nu}(Z')}{g^2_{\rm SM}}\right]
\frac{\bar\Delta_{L}^{sb}(Z')}{ V_{ts}^\ast V_{tb}}\equiv \eta_X X_0(x_t)+C_{\nu\bar\nu}^{\rm NP},
\ee
where the first term is the SM contribution given in  \cite{Buras:2012jb}.
Using the relevant formulae in \cite{Altmannshofer:2009ma}, unique predictions, 
independent of the sign of  $\bar\Delta_L^{\nu\bar\nu}(Z')$, for $B_d\to K\nu\bar\nu$, $B_d\to K^*\nu\bar\nu$ and $B_d\to X_s\nu\bar\nu$ can be made. Indeed 
we obtain 
by means of (\ref{SU2}) an important relation
\be\label{nunu910}
C_{\nu\bar\nu}^{\rm NP}=\sin^2\theta_W\frac{C_{10}^{\rm NP}-C_9^{\rm NP}}{2}.
\ee

\section{Correlations between flavour observables in the LHS}\label{sec:4}
\subsection{Preliminaries}
Having all these results at hand we want to illustrate this procedure 
and the implied correlations numerically. As already pointed out and 
analyzed in detail  in 
\cite{Buras:2012jb} due the strong correlations between $\Delta F=2$ and 
$\Delta F=1$ observables within LHS, the pattern and size of allowed
 NP effects in the latter 
observables depends crucially on the room left for NP contributions in 
$\Delta F=2$ observables. Therefore we briefly review the situation in 
$B_{s,d}-\bar B_{s,d}$ mixings. In contrast to CMFV models we do not have 
to discuss simultaneously the $\Delta M_K$ and $\varepsilon_K$ 
as generally there is no correlation of these observables with $B_{s,d}-\bar B_{s,d}$ mixings. This allows to avoid certain tension in CMFV models 
analysed recently by us \cite{Buras:2013raa}.

It is useful to recall the parametric dependence of $\Delta M_{s,d}$ 
within the SM.  One has\footnote{The central value of $\vts$ corresponds roughly to the central $\vcb=0.0409$ obtained from tree-level
decays.}
\begin{equation}\label{DMS}
(\Delta M_{s})_{\rm SM}=
17.7/{\rm ps}\cdot\left[ 
\frac{\sqrt{\hat B_{B_s}}F_{B_s}}{267\mev}\right]^2
\left[\frac{S_0(x_t)}{2.31}\right] 
\left[\frac{\vts}{0.0402} \right]^2
\left[\frac{\eta_B}{0.55}\right] \,,
\end{equation}

\begin{equation}\label{DMD}
(\Delta M_d)_{\rm SM}=
0.51/{\rm ps}\cdot\left[ 
\frac{\sqrt{\hat B_{B_d}}F_{B_d}}{218\mev}\right]^2
\left[\frac{S_0(x_t)}{2.31}\right] 
\left[\frac{\vtd}{8.5\cdot10^{-3}} \right]^2 
\left[\frac{\eta_B}{0.55}\right]
\end{equation}
We observe that for the chosen central values of the parameters there 
is a perfect agreement with the very accurate data \cite{Amhis:2012bh}:
\be\label{exp}
\Delta M_s = 17.69(8)\,\text{ps}^{-1}, \qquad \Delta M_d = 0.510(4)\,\text{ps}^{-1}~.
\ee

In fact these central values are very close to the central values presented 
recently by the Twisted Mass Collaboration \cite{Carrasco:2013zta}
\be\label{twist}
\sqrt{\hat B_{B_s}}F_{B_s} = 262(10)\mev,\qquad \sqrt{\hat B_{B_d}}F_{B_d} = 216(10)\mev.
\ee
Similar values $266(18)\mev$ and $216(15)$ with larger errors are quoted 
by FLAG, where the new results in (\ref{twist}) are not yet included.
While the central values in (\ref{twist}) are lower than the ones used by us 
in \cite{Buras:2012jb} they agree with the latter within the errors, in particular when the numbers from FLAG are considered.

Concerning $S_{\psi\phi}$ and $S_{\psi K_S}$ we have
\be
S_{\psi\phi}= -\left(0.04^{+0.10}_{-0.13}\right), \qquad S_{\psi K_S}= 0.679(20)
\ee
with the second value known already for some time \cite{Amhis:2012bh} and the 
first one being the most recent average from HFAG \cite{Amhis:2012bh} close 
to the earlier result from the LHCb \cite{Aaij:2013oba}. 
The first value is consistent with the SM expectation of $0.04$.  This is also 
the case of  $S_{\psi K_S}$ but only for exclusive determination of $\vub\approx
(3.4\pm0.2)\times 10^{-3}$  \cite{Buras:2013ooa}. When the inclusive determination of $\vub$, like $(4.2\pm0.2)\times 10^{-3}$, 
  is used $S_{\psi K_S}$ is found above $0.77$
and 
NP is required to bring it down to its experimental value with 
interesting implications for $\Delta F=1$ observables as shown in 
\cite{Buras:2012jb} and below.

While there is some feeling in the community that there is no NP in $S_{\psi\phi}$ we would like to stress again \cite{Buras:2013ooa} that a precise measurement of this observable is very important as it provides powerful tests of various NP scenarios. In particular the distinction between CMFV models based on 
$U(3)^3$ flavour symmetry and $U(2)^3$ models can be performed by studying 
triple correlation  $S_{\psi\phi}-S_{\psi K_S}-\vub$ \cite{Buras:2012sd}. 
 We will also see below that this determination is also important for the 
tests of the  LHS model  through the observables considered here.

\subsection{Strategy}
It is to be expected that in flavour precision era, which will include both 
advances in experiment and theory, in particular lattice calculations, 
it will be possible to decide with high precision 
whether $\Delta M_s$ and $\Delta M_d$ within the SM agree or disagree with 
the data. For instance already the need for enhancements or suppressions 
of these observables would be an important information. Similar comments 
apply to $S_{\psi\phi}$ and $S_{\psi K_S}$ as well as to the branching 
ratios $\mathcal{B}(B_{s,d}\to\mu^+\mu^-)$. In particular the correlations 
and anti-correlations between the suppressions and enhancements allow 
to distinguish between various NP models as can be illustrated with the DNA 
charts proposed in  \cite{Buras:2013ooa}.

In order to monitor this progress 
in the context of the LHS model  we use the ratios \cite{Bona:2006sa}:
\be
C_{B_s}=\frac{\Delta M_s}{(\Delta M_s)_{\rm SM}}=\frac{|S(B_s)|}{S_0(x_t)}, \qquad C_{B_d}=\frac{\Delta M_d}{(\Delta M_d)_{\rm SM}}=\frac{|S(B_d)|}{S_0(x_t)}
\ee
that in the LHS, thanks to the presence of a single operator do not involve any non-perturbative uncertainties. Of course in order 
to find out the experimental values of these ratios one has to handle these 
uncertainties but this is precisely what we want to monitor in the coming 
years.  The most recent update from Utfit collaboration reads
\be\label{UTfit}
C_{B_s}=1.08\pm 0.09, \qquad C_{B_d}=1.10\pm 0.17~.
\ee

\subsection{Performing steps 1 and 2}
\begin{figure}[!tb]
 \centering
\includegraphics[width = 0.47\textwidth]{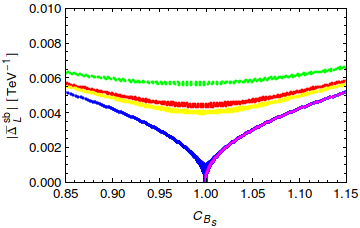}
\includegraphics[width = 0.45\textwidth]{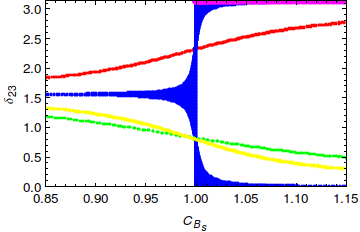}
\caption{$\left|\bar\Delta_L^{sb}\right|$ and $\delta_{23}$ versus $C_{B_s} = \Delta M_s/\Delta 
M_s^\text{SM}$ for $S_{\psi\phi}\in[0.14,0.15]$ (red), [0.03,0.04] (blue), [-0.06,-0.05] (yellow) and 
[-0.15,-0.14] (green). The magenta line is for real $C_9^\text{NP}$, i.e. for $\delta_{23} = \beta_s (+\pi)$. 
}\label{fig:BsDeltaF2}~\\[-2mm]\hrule
\end{figure}

\begin{figure}[!tb]
 \centering
\includegraphics[width = 0.47\textwidth]{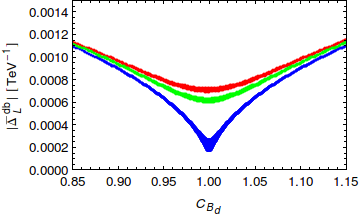}
\includegraphics[width = 0.45\textwidth]{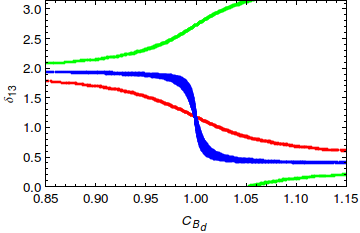}
\caption{$\left|\bar\Delta_L^{db}\right|$ and $\delta_{13}$ versus $C_{B_d} = \Delta M_d/\Delta 
M_d^\text{SM}$ for $\vub = 0.0034$ and  $S_{\psi K_S}\in[0.718,0.722]$ (red), [0.678,0.682] (blue), [0.638,0.642] (green). 
}\label{fig:BdDeltaF2lowVub}~\\[-2mm]\hrule
\end{figure}

\begin{figure}[!tb]
 \centering
\includegraphics[width = 0.47\textwidth]{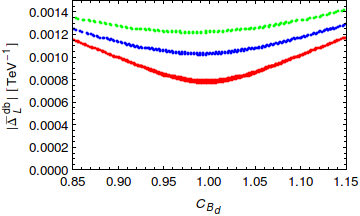}
\includegraphics[width = 0.45\textwidth]{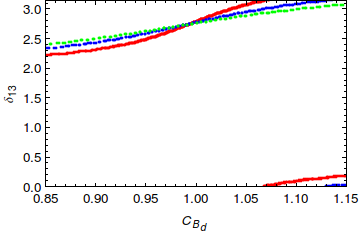}
\caption{$\left|\bar\Delta_L^{db}\right|$ and $\delta_{13}$ versus $C_{B_d} = \Delta M_d/\Delta 
M_d^\text{SM}$ for $\vub = 0.0040$ and  $S_{\psi K_S}\in[0.718,0.722]$ (red), [0.678,0.682] (blue), [0.638,0.642] (green). 
}\label{fig:BdDeltaF2highVub}~\\[-2mm]\hrule
\end{figure}

Using this strategy we can perform Step 1. In Fig.~\ref{fig:BsDeltaF2} we show in the case 
of the $B_s^0-\bar B_s^0$ system $|\bar\Delta_L^{sb}(Z')|$ and the phase $\delta_{23}$ as functions of $C_{B_s}$ for different values of $S_{\psi\phi}$. 
The results are given for the blue oasis (low $\delta_{23}$) . For the purple oasis  (high $\delta_{23}$) one just has 
to add $180^\circ$ to $\delta_{23}$. In Fig.~\ref{fig:BdDeltaF2lowVub} we repeat this exercise for the 
 $B_d^0-\bar B_d^0$ system setting $\beta=21.2^\circ$ extracted using the exclusive value 
of $\vub=3.4\times 10^{-3}$. This corresponds to the LHS1 scenario 
of \cite{Buras:2012jb}. In Fig.~\ref{fig:BdDeltaF2highVub} we show the results for LHS2 scenario 
 with $\beta=25.2^\circ$ corresponding to $\vub=4.0\times 10^{-3}$ and in the 
ballpark of inclusive determinations of this CKM element.

Our colour coding for the values of $S_{\psi\phi}$ and $S_{\psi K_S}$ 
is as follows:
\begin{itemize}
\item
In the case of  $B_s^0-\bar B_s^0$ we fix $S_{\psi\phi}$ to the following 
ranges: {\it red} ($0.14-0.15$), {\it blue} $(0.03-0.04)$, {\it yellow} $(-(0.05-0.06)$), {\it green} $(-(0.14-0.15))$.
\item
In the case of  $B_d^0-\bar B_d^0$ we fix $S_{\psi K_S}$ to the following 
ranges: {\it red} ($0.718-0.722$), {\it blue} $(0.678-0.682)$, {\it green} $(0.638-0.642)$.
\end{itemize}

\begin{figure}[!tb]
 \centering
\includegraphics[width = 0.5\textwidth]{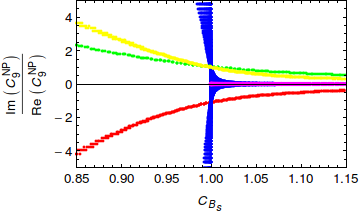}
\caption{$\IM(C_9^\text{NP})/\RE(C_9^\text{NP}) = \IM(C_{10}^\text{NP})/\RE(C_{10}^\text{NP})$ versus $C_{B_s} = \Delta 
M_s/\Delta M_s^\text{SM}$   for $S_{\psi\phi}\in[0.14,0.15]$ (red), [0.03,0.04] (blue), [-0.06,-0.05] (yellow) and 
[-0.15,-0.14] (green). The magenta line is for real $C_9^\text{NP}$, i.e. for $\delta_{23} = \beta_s (+\pi)$.
}\label{fig:Bstan}~\\[-2mm]\hrule
\end{figure}

The following observations can be made on the basis of these plots.
\begin{itemize}
\item
The value of  $|\bar\Delta_L^{sb}(Z')|$ vanishes in the blue scenario when 
$C_{B_s}=1$ as in this scenario also $S_{\psi\phi}$ is very close to its 
SM value. In other scenarios for $S_{\psi\phi}$ the coupling $|\bar\Delta_L^{sb}(Z')|$ is non-vanishing even 
for $C_{B_s}=1$ as LHS has to bring  $S_{\psi\phi}$ to agree with data. 
The more the value of  $S_{\psi\phi}$ differs from the SM one, the larger 
 $|\bar\Delta_L^{sb}(Z')|$ is allowed to be. Its value increases with increasing
  $|C_{B_s}-1|$ and this increase is strongest in the blue scenario.
\item
The value of $\delta_{23}$ in the blue scenario is arbitrary for 
$C_{B_s}=1$ but as in this case  $|\bar\Delta_L^{sb}(Z')|$ vanishes 
NP contributions vanish anyway. In this scenario for $C_{B_s}< 1$ 
we find $\delta_{23}\approx \pi/2$. For $C_{B_s}>1$  one has 
 $\delta_{23}\approx (0,\pi)$. In the red scenario, in which $S_{\psi\phi}$ is 
larger than its SM value  $\delta_{23}$ is confined 
to the second quadrant and increases monotonically with increasing 
$C_{B_s}$. In the yellow and green scenario in which $S_{\psi\phi}$ is 
negative, $\delta_{23}$ is confined to the first quadrant and decreases 
monotonically with increasing $C_{B_s}$. 

\item
We note that in red, yellow 
and green scenarios for $S_{\psi\phi}$, in
the full range of $C_{B_s}$ considered, $\delta_{23}$ has to differ from 
$0$, $\pi/2$ or $\pi$. Through 
(\ref{IMRE}) this implies that the coefficients 
$C_9^{\rm NP}$ and $C_{10}^{\rm NP}$ cannot be real. We show this in Fig.~\ref{fig:Bstan} 
for these three scenarios of $S_{\psi\phi}$. 
We observe that for green and yellow scenarios for which $S_{\psi\phi}<0$, 
the ratio in question is positive, while it is negative for  $S_{\psi\phi}$ 
larger than the SM value. Moreover for $C_{B_s}$ significantly smaller 
than unity, where large NP phase is required to obtain correct value for 
$\Delta M_s$  the ratios in (\ref{IMRE}) have
 to be large. This is not the 
case for $C_{B_s}>1$ as then  a real $C_9^{\rm NP}$ implies automatically 
an enhancement of $C_{B_s}$ as we will discuss in Section~\ref{sec:5}. 

\item
Turning now our discussion to the $B_d^0-\bar B_d^0$ system we observe that 
in the LHS1 scenario $\delta_{13}$ is confined to the first quadrant in the 
red scenario where $S_{\psi K_S}$ is larger than its SM value and dominantly 
in the second quadrant (green scenario) when it is below it. In the LHS2 scenario 
$|\bar\Delta_L^{db}(Z')|$ is always different from zero as the SM value of 
$S_{\psi K_S}$ being in our example $0.77$ is larger than its experimental value in red, blue and green  scenarios considered and NP has to remove this discrepancy.
\end{itemize}

In summary once $S_{\psi\phi}$, $C_{B_s}$, $S_{\psi K_S}$ and $C_{B_d}$ are precisely known plots in Figs.~\ref{fig:BsDeltaF2}-\ref{fig:BdDeltaF2highVub} will allow to obtain 
precise values of $|\bar\Delta_L^{sb}(Z')|$, $\delta_{23}$,  $|\bar\Delta_L^{db}(Z')|$, $\delta_{13}$ allowing the predictions for other observables.

\subsection{Performing step 3} 
In \cite{Buras:2012jb} we have set $\bar\Delta_A^{\mu\bar\mu}(Z')=0.5/\tev$.
The goal of this step is to determine which values of this coupling are 
consistent with the present values of the branching ratios for 
$B_{s,d}\to \mu^+\mu^-$ decays. The status of these branching ratios has been 
summarized in~(\ref{LHCb2}) and~(\ref{LHCb3}).

In order to get an idea what values of  $\bar\Delta_A^{\mu\bar\mu}(Z')$ 
are consistent with these data and with the data on CP-asymmetries 
$S_{\psi\phi}$ and $S_{\psi K_S}$, we study the correlations 
between $\overline{\mathcal{B}}(B_{s}\to\mu^+\mu^-)$ and $S_{\psi\phi}$ and 
between $\mathcal{B}(B_{d}\to\mu^+\mu^-)$ and $S_{\psi K_S}$ for four different values of  $|\bar\Delta_A^{\mu\bar\mu}(Z')|$ in five bins for $C_{B_s}$ and 
$C_{B_d}$ all to be listed below. This time  in order to have a
 comparison 
with \cite{Buras:2012jb} we present the results for two oases:  low and 
high $\delta_{23}$ for $B_s\to \mu^+\mu^-$ and  low and 
high $\delta_{13}$ for $B_d\to \mu^+\mu^-$. Yet in order not to complicate 
colour coding  we use the same colours for the two oases.
The main difference with respect to \cite{Buras:2012jb} is the study 
of the dependence on  $|\bar\Delta_A^{\mu\bar\mu}(Z')|$ and
the consideration 
of values of $C_{B_{s,d}}$ smaller and larger than unity, whereas in 
that paper only $C_{B_{s,d}}< 1.0$ have been considered due to previous 
lattice results as discussed above. Moreover, anticipating progress in 
lattice calculations we consider several bins in $C_{B_{s,d}}$. 

\begin{figure}[!tb]
 \centering
\includegraphics[width = 0.45\textwidth]{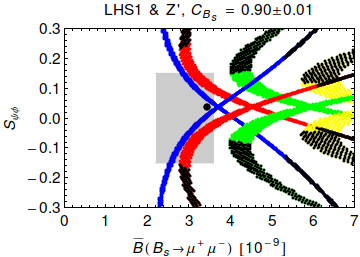}
\includegraphics[width = 0.45\textwidth]{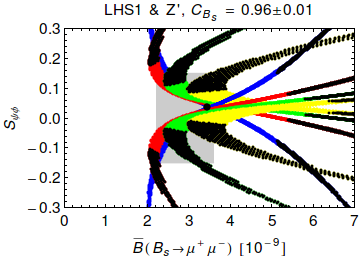}
\includegraphics[width = 0.45\textwidth]{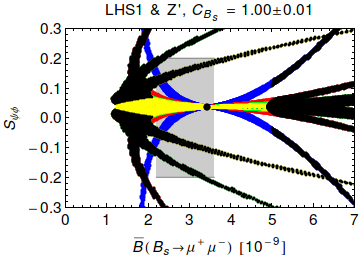}
\includegraphics[width = 0.45\textwidth]{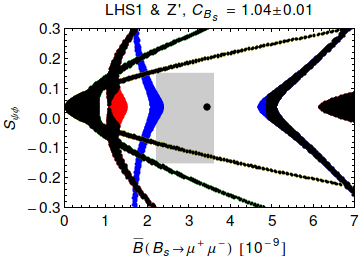}
\begin{flushleft}
 \hspace{0.6cm}
\includegraphics[width = 0.45\textwidth]{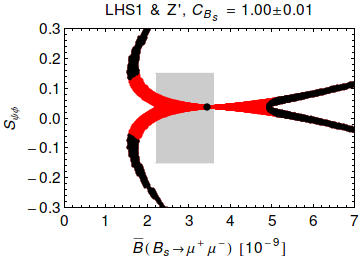}
\end{flushleft}

\caption{$S_{\psi\phi}$ versus $\overline{\mathcal{B}}(B_{s}\to\mu^+\mu^-)$ for different values of $C_{B_s}$ 
(see Eq.~(\ref{bins})) and $\bar\Delta_A^{\mu\bar\mu}$ (see Eq.~(\ref{DA})).  In the last plot only $\bar\Delta_A^{\mu\bar\mu} =
1~\text{TeV}^{-1}$ for $C_{B_s} = 1.00\pm0.01$ is shown. The black  regions violate the bounds 
for $C_{10}^\text{NP}$ (see Eq.~(\ref{equ:C10NPconstraints})). Black point: SM central value. The gray region represents the data.}\label{fig:binBs}~\\[-2mm]\hrule
\end{figure}

\begin{figure}[!tb]
 \centering
\includegraphics[width = 0.45\textwidth]{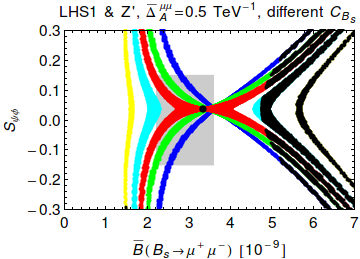}
\includegraphics[width = 0.45\textwidth]{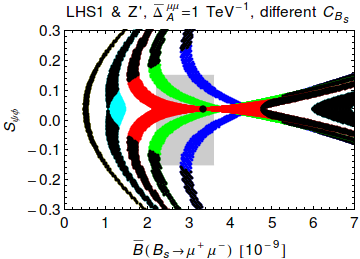}
\includegraphics[width = 0.45\textwidth]{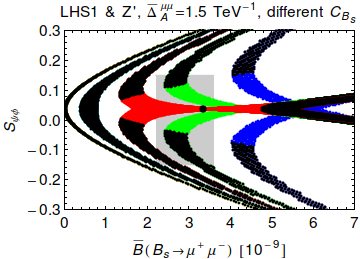}
\includegraphics[width = 0.45\textwidth]{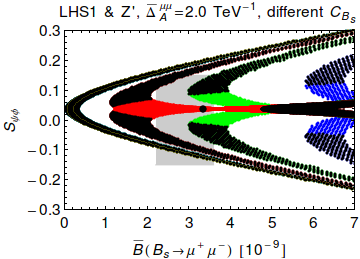}
\caption{$S_{\psi\phi}$ versus $\overline{\mathcal{B}}(B_{s}\to\mu^+\mu^-)$ for different values of $C_{B_s}$ 
(see Eq.~(\ref{bins})) and fixed $\bar\Delta_A^{\mu\bar\mu}$ (see caption of the plot).  The black  regions violate the bounds 
for $C_{10}^\text{NP}$ (see Eq.~(\ref{equ:C10NPconstraints})). Black point: SM central value. The gray region represents the
data.}\label{fig:binBsneu}~\\[-2mm]\hrule
\end{figure}

\begin{figure}[!tb]
 \centering
\includegraphics[width = 0.45\textwidth]{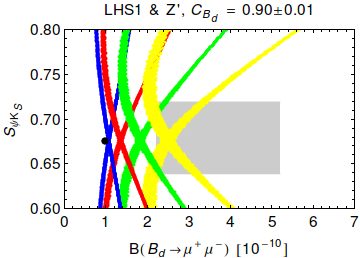}
\includegraphics[width = 0.45\textwidth]{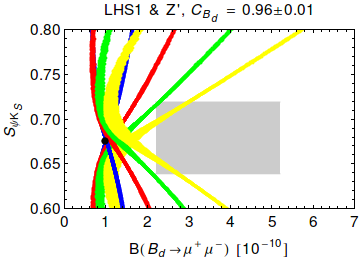}
\includegraphics[width = 0.45\textwidth]{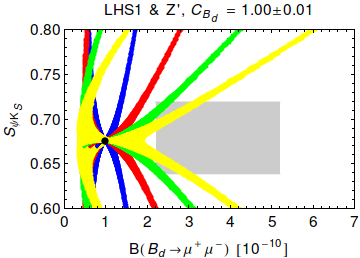}
\includegraphics[width = 0.45\textwidth]{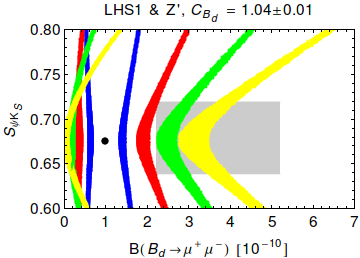}
\caption{$S_{\psi K_S}$ versus ${\mathcal{B}}(B_{d}\to\mu^+\mu^-)$ for $\vub = 0.0034$ and different values of $C_{B_d}$ 
(see Eq.~(\ref{bins})) and $\bar\Delta_A^{\mu\bar\mu}$ (see Eq.~(\ref{DA})).  Black point: SM central
value. The gray region represents the data.}\label{fig:binBdlowVub}~\\[-2mm]\hrule
\end{figure}

\begin{figure}[!tb]
 \centering
\includegraphics[width = 0.45\textwidth]{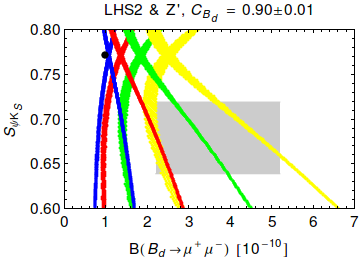}
\includegraphics[width = 0.45\textwidth]{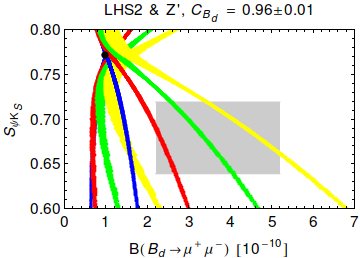}
\includegraphics[width = 0.45\textwidth]{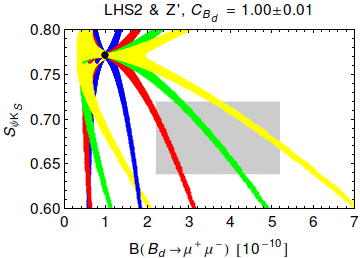}
\includegraphics[width = 0.45\textwidth]{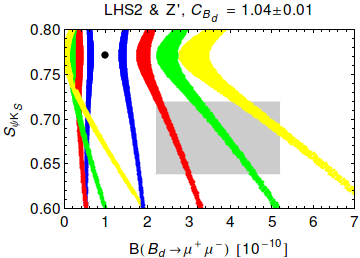}
\caption{$S_{\psi K_S}$ versus ${\mathcal{B}}(B_{d}\to\mu^+\mu^-)$ for $\vub = 0.0040$ and different values of $C_{B_d}$ 
(see Eq.~(\ref{bins})) and $\bar\Delta_A^{\mu\bar\mu}$ (see Eq.~(\ref{DA})).  Black point: SM central
value.  The gray region represents the data.}\label{fig:binBdhighVub}~\\[-2mm]\hrule
\end{figure}

It turns out that the results for the correlations of $\overline{\mathcal{B}}(B_{s}\to\mu^+\mu^-)$ and $S_{\psi\phi}$ look rather involved and in order to increase the transparency of our presentation we have in this case shown two figures. 
In Fig.~\ref{fig:binBs} we show five plots each corresponding to different bin 
in $C_{B_s}$ and in each plot different colours correspond to different values 
of $\bar\Delta_A^{\mu\bar\mu}(Z')$. On the other hand in Fig.~\ref{fig:binBsneu} 
we show four plots corresponding to different values of $\bar\Delta_A^{\mu\bar\mu}(Z')$ and in each plot different colours correspond to different values of 
$C_{B_s}$. The latter presentation turns out to be unnecessary in the case 
of the correlation  $\mathcal{B}(B_{d}\to\mu^+\mu^-)$ and $S_{\psi K_S}$ for 
which the results are presented in Figs.~\ref{fig:binBdlowVub} and \ref{fig:binBdhighVub} for the $B_d$ system 
in LHS1 and LHS2 scenarios for $\vub$, respectively.

The colour coding for  $\bar\Delta_A^{\mu\bar\mu}(Z')$ in both $B_s$ and $B_d$ 
systems is
\be\label{DA}
\bar\Delta_A^{\mu\bar\mu}(Z')= 0.50/\tev~\text{(blue)},\quad 1.00/\tev~\text{(red)},\quad 1.50/\tev~\text{(green)}, \quad 2.00/\tev~\text{(yellow)}.
\ee
Working with two oases it is sufficient to consider only positive values of  $\bar\Delta_A^{\mu\bar\mu}(Z')$
as discussed above.

On the other hand the colour coding for $C_{B_s}$ relevant only for 
Fig.~\ref{fig:binBsneu} is as follows
\begin{align}\begin{split}\label{bins}
& C_{B_s}=0.90\pm0.01~\text{(blue)}, \quad 0.96\pm 0.01~\text{(green)},\quad 1.00\pm 0.01~\text{(red)}, \\&
C_{B_s}= 1.04\pm 0.01~\text{(cyan)}, \quad 1.10\pm 0.01~\text{(yellow)}\end{split}
\end{align}
and similarly for $C_{B_d}$. 
Furthermore we include bounds for $C_{10}^\text{NP}$ derived from $b\to s\ell\ell$ transitions 
\cite{Altmannshofer:2012ir,Straub:2013uoa,Altmannshofer:2013oia}. As in \cite{Buras:2013ooa} we use the following range 
\begin{align}\label{equ:C10NPconstraints}
&-0.8\leq \RE(C_{10}^\text{NP})\leq 1.8\,,\quad -3\leq \IM(C_{10})\leq 3\,.
\end{align}
The exact bounds are even smaller than these rectangular bounds.  Clearly
these bounds will be changing with time together with the modification of 
data but we show their impact here to illustrate how looking simultaneously 
at various observables can constrain a given NP scenario.

The main results of this exercise 
are as follows.
\begin{itemize}
\item
In the case of the $B_s$ system the  correlation between 
$\overline{\mathcal{B}}(B_{s}\to\mu^+\mu^-)$ and $S_{\psi\phi}$ 
and the allowed values of  $\bar\Delta_A^{\mu\bar\mu}(Z')$ depend 
strongly on the chosen value of $C_{B_s}$.  Moreover, the constraints 
in (\ref{equ:C10NPconstraints}), that eliminate  the regions in black,
have an important impact on these plots. The larger $C_{B_s}$, the more 
points are eliminated. For the last bin $C_{B_s}=1.10\pm 0.01$ the only 
points that survive are the  ones with $\bar\Delta_A^{\mu\bar\mu}= 0.50/\tev$
and $\overline{\mathcal{B}}(B_{s}\to\mu^+\mu^-)\le 1.7\cdot 10^{-9}$ so that 
this bin is excluded. This can be seen in Fig.~\ref{fig:binBsneu} and 
therefore we do not show this bin in  Fig.~\ref{fig:binBs}
\item
As seen in Fig.~\ref{fig:binBsneu} for  $C_{B_s}=0.96\pm0.01$ (green)
basically all values of $\bar\Delta_A^{\mu\bar\mu}(Z')$ in (\ref{DA}) are 
allowed but this is no longer the case for other bins of  $C_{B_s}$. 
In particular for  $C_{B_s}\approx 0.90$ (blue) as considered in \cite{Buras:2012jb} the values $\bar\Delta_A^{\mu\bar\mu}(Z')\le 1.0$ are favoured. In this case 
not only present data on $\overline{\mathcal{B}}(B_{s}\to\mu^+\mu^-)$ and $S_{\psi\phi}$  can be reproduced but also values close to SM expectations for 
these two observables are allowed.
\item
On the other hand  for $C_{B_s}= 1.04$ (cyan) only $\bar\Delta_A^{\mu\bar\mu}(Z')\approx 0.5$ is allowed by all constraints and $\overline{\mathcal{B}}(B_{s}\to\mu^+\mu^-)\le 2.5\cdot 10^{-9}$ or around $5\cdot 10^{-9}$. That is for this value 
of $C_{B_s}= 1.04$ the branching ratio in question differs sizably from the 
SM predictions and in fact $\bar\Delta_A^{\mu\bar\mu}(Z')< 0.5$ would 
give a better agreement with the data. But as we discuss below one 
needs $\bar\Delta_A^{\mu\bar\mu}(Z')\ge 1.0$ to obtain the agreement 
with the data for $\mathcal{B}(B_{d}\to\mu^+\mu^-)$.  We conclude therefore 
that the present data on $B_{s,d}\to\mu^+\mu^-$ when the constraints in 
 (\ref{equ:C10NPconstraints}) are taken into account favour $C_{B_s}\le 1.0$. 
Thus 
$\bar\Delta_A^{\mu\bar\mu}(Z')=1.0$ and  $C_{B_s}= 1.00\pm0.01$ appears 
to us to be the optimal combination when also $B_d\to\mu^+\mu^-$ is considered.
\item
Indeed moving next to the $B_d$ system we observe that basically only for
 $\bar\Delta_A^{\mu\bar\mu}(Z')\ge 1.0$ can one reproduce 
the data for  $\mathcal{B}(B_{d}\to\mu^+\mu^-)$ within $1\sigma$. Moreover, 
it is striking that measuring this branching ratio and  $S_{\psi K_S}$ precisely 
it is much easier to determine the coupling $\bar\Delta_A^{\mu\bar\mu}(Z')$ than 
through the $B_s$ system. This is of course the consequence of the large NP 
contribution needed to come close to the central values of the 
CMS and LHCb data.
We also observe 
that NP effects are larger in LHS2 scenario as in this case there is a bigger 
room for NP in $S_{\psi K_s}$ which has to be suppressed to agree with the data.
\item
We observe that if one does not want to work with values of  $\bar\Delta_A^{\mu\bar\mu}(Z')$ as high as  $1.5/\tev$  and $2.0/\tev$ 
the data on $B_d\to\mu^+\mu^-$ favour $C_{B_d}\ge 1.00$ and LHS2 scenario, although for $C_{B_d}\ge 1.04$ and $\bar\Delta_A^{\mu\bar\mu}(Z')=1.0/\tev$ also 
interesting results in LHS1 are obtained.
\end{itemize}

To summarize, the present data allow to find certain pattern in the couplings:
\begin{itemize}
\item
The $B_s$ system favours $C_{B_s}\le 1.00$, while the $B_d$-system 
$C_{B_d}\ge 1.00$ . Yet these two values cannot differ by more than $5\%$ 
in order to reproduce the experimental ratio $\Delta M_s/\Delta M_d$ which 
contains smaller hadronic uncertainties than $\Delta M_s$ and $\Delta M_d$ 
separately.
\item
It appears that  $\bar\Delta_A^{\mu\bar\mu}(Z')\approx 1.0/\tev$ is the 
present favoured value for this coupling but it will go down if 
the unexpectedly large values of $\mathcal{B}(B_{d}\to\mu^+\mu^-)$ will 
decrease in the future.
\end{itemize}

With this information at hand let us see  what impact these  results have 
on the correlation between the two branching ratios in question.
As Figs.~\ref{fig:binBs}-\ref{fig:binBdhighVub} allow to find out what is the outcome of this exercise
 we only show two examples by setting 
\be
 C_{B_s}=1.00\pm0.01, \qquad C_{B_d}=1.04\pm 0.01, \qquad \bar\Delta_A^{\mu\bar\mu}(Z')=1.0/\tev
\ee
and calculating $\mathcal{B}(B_{d}\to\mu^+\mu^-)$ for the cases of LHS1 and LHS2. We varied the CP-asymmetries in the following ranges
\be\label{rangeS}
-0.15\le S_{\psi\phi}\le 0.15, \qquad 0.639\le S_{\psi K_S}\le 0.719\,.
\ee

\begin{figure}[!tb]
 \centering
\includegraphics[width = 0.45\textwidth]{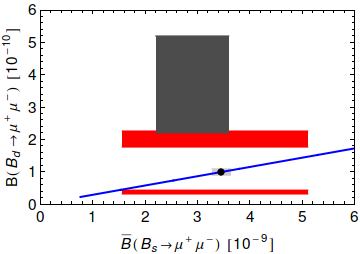}
\includegraphics[width = 0.45\textwidth]{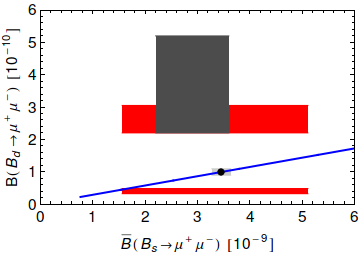}
\caption{ ${\mathcal{B}}(B_{d}\to\mu^+\mu^-)$ versus ${\mathcal{\bar{B}}}(B_{s}\to\mu^+\mu^-)$ for $\vub = 0.0034$ (left) and $\vub =
0.0040$ (right) and  $C_{B_d} = 1.04\pm 0.01$, $C_{B_s} = 1.00\pm 0.01$, $\bar{\Delta}_A^{\mu\bar\mu} = 1~\text{TeV}^{-1}$, $0.639\leq
S_{\psi K_s}\leq 0.719$ and $-0.15\leq S_{\psi\phi}\leq 0.15$. SM is represented by the light gray area with black dot and 
 the CMFV prediction by the blue line. Dark gray
 region: Combined exp 1$\sigma$
 range
 $\overline{\mathcal{B}}(B_s\to\mu^+\mu^-) = (2.9\pm0.7)\cdot 10^{-9}$ and $\mathcal{B}(B_d\to\mu^+\mu^-) = (3.6^{+1.6}_{-1.4})\cdot
10^{-10}$.}\label{fig:BdvsBsLHS}~\\[-2mm]\hrule
\end{figure}

The result is shown in Fig.~\ref{fig:BdvsBsLHS} where the blue line corresponds  to $r=1$ in (\ref{CMFV6}). For both LHS1 and LHS2 
there are two regions corresponding to enhanced and suppressed values 
of  $\mathcal{B}(B_{d}\to\mu^+\mu^-)$. In the LHS2 case one finds that these 
two regions correspond to two different oases. Similar structure specific 
to the $C_{B_d}> 1$ region and LHS2 has been found in 
the context of the analysis of 331 models in \cite{Buras:2012dp} (see Fig. 20 of that paper). The origin of this behaviour is explained in detail in that 
paper with large values of  $\mathcal{B}(B_{d}\to\mu^+\mu^-)$ corresponding 
to larger phase $\delta_{13}$ for a chosen positive sign of  $\bar\Delta_A^{\mu\bar\mu}(Z')$. We find that in LHS1 case enhancement of  $\mathcal{B}(B_{d}\to\mu^+\mu^-)$ corresponds for $S_{\psi K_S}\ge 0.66$ to low $\delta_{13}$ values but 
for  $S_{\psi K_S}\le 0.66$ both high and low ranges for $\delta_{13}$ can 
provide the enhancement. Note that in the case of enhancement 
 $\mathcal{B}(B_{d}\to\mu^+\mu^-)$ overlaps with the data, while in the case 
of suppression it is roughly  by a factor of two to three below its SM value.
We should mention that we assume here that the two measurements of $B_s\to\mu^+\mu^-$ and $B_d\to\mu^+\mu^-$ are uncorrelated which is
however not the case. This is shown for example in Fig.~2 of \cite{Chatrchyan:2013bka}. Consequently the areas in our
Fig.~\ref{fig:BdvsBsLHS} are exactly rectangular which would change if a correlation matrix for the combined LHCb and CMS data was
included. This difference will only matter when the data improve. 

Analogous structure would be found in $B_s\to \mu^+\mu^-$ case if we had chosen 
 $C_{B_s}=1.04$, but in this case as already noticed in \cite{Buras:2012dp} 
(see Fig.~21 of that paper) and also seen in the corresponding plot in 
Fig.~\ref{fig:binBs} of the present paper, the impact of moving from 
$C_{B_s}<1.0$ to $C_{B_s}>1.0$ is much larger than in the $B_d$ case and for 
$\bar\Delta_A^{\mu\bar\mu}(Z')\approx 1.0$ as required to fit the present 
data on  $\mathcal{B}(B_{d}\to\mu^+\mu^-)$ one fails in this case to fit data 
for
$\overline{\mathcal{B}}(B_{s}\to\mu^+\mu^-)$. In fact  the black area 
 most to the right in 
Fig.~\ref{fig:binBs} for $C_{B_s}=1.04$  would be red if we did not use the constraints~(\ref{equ:C10NPconstraints}). Also the blue region
in this plot corresponding to 
$\bar\Delta_A^{\mu\bar\mu}(Z')=0.5$ shows this structure.

For our choice of $C_{B_s}=1.00$  the effects in question are mixed up 
in the two oases and there is no clear correspondence between a given 
oasis and enhancement or suppression of $\overline{\mathcal{B}}(B_{s}\to\mu^+\mu^-)$ when $S_{\psi\phi}$  is varied.

We concentrate now on the red regions in which $\mathcal{B}(B_{d}\to\mu^+\mu^-)$ is close to the data.
Inspecting previous plots we find that the asymmetries $S_{\psi\phi}$ and 
$S_{\psi K_S}$ serve as coordinates in the horizontal and vertical direction, 
respectively and the departure from the SM point increases with the 
increasing $|S_{\psi K_S}-S_{\psi K_S}^{\rm NP}|$ and similarly for the horizontal 
direction. Therefore
\begin{itemize}
\item
In the LHS1 case the largest  $\mathcal{B}(B_{d}\to\mu^+\mu^-)$ is obtained 
for the largest and lowest value of  $S_{\psi K_S}$ in (\ref{rangeS}).
\item
In the LHS2 case it is found for the lowest value of $S_{\psi K_S}$.
\item 
For LHS1 and LHS2 we find respectively
\be 
1.8\cdot 10^{-10}\le\mathcal{B}(B_{d}\to\mu^+\mu^-)\le 2.3\cdot 10^{-10}, \qquad 
{(\rm LHS1)}.
\ee

\be 
2.2\cdot 10^{-10}\le\mathcal{B}(B_{d}\to\mu^+\mu^-)\le 3.1\cdot 10^{-10}, \qquad
{(\rm LHS2)}.
\ee
\end{itemize}

 We also observe that in LHS2 even for $\bar\Delta_A^{\mu\bar\mu}(Z')=0.5/\tev$ 
(see Fig.~\ref{fig:binBdhighVub}) this branching ratio can be by a factor of $1.5-1.9$ larger than 
its SM value when  $C_{B_d}=1.04$.

Inspecting the plots in Figs.~\ref{fig:binBs}-\ref{fig:binBdhighVub} we 
also observe that for $C_{B_d}=1.00$ there would be no separation in two 
oases in the case of LHS1 and the values of $\mathcal{B}(B_{d}\to\mu^+\mu^-)$ 
would be lower than in the example presented in Fig.~\ref{fig:BdvsBsLHS} but 
the plot for LHS2 would hardly change. With decreasing  $C_{B_d}$ the LHS1 
scenario is unable to reproduce the data for the branching ratio in question 
for $\bar\Delta_A^{\mu\bar\mu}(Z')=1.0/\tev$, while in the case of LHS2 the data
for this branching ratio within $1\sigma$ can be reproduced provided $S_{\psi K_S}$ is low enough.

\boldmath
\subsection{The $b\to s \nu\bar\nu$ Transitions}
\unboldmath

\begin{figure}[!tb]
 \centering
\includegraphics[width = 0.5\textwidth]{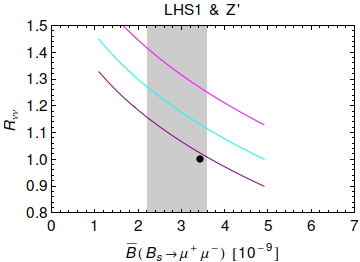}
\caption{ $R_{\nu\bar\nu}$ versus $\overline{\mathcal{B}}(B_s\to\mu^+\mu^-)$ for  $C_9^\text{NP} = -1.6$ (magenta),
$C_9^\text{NP} =
-0.8$ (cyan), $C_9^\text{NP} =
-0.14$ (purple) and $-0.8\leq C_{10}^\text{NP}\leq 1.8$.
}\label{fig:BsmuRnu}~\\[-2mm]\hrule
\end{figure}

In the absence of right-handed currents one finds \cite{Altmannshofer:2009ma}
\be
R_{\nu\bar\nu}\equiv
 \frac{\mathcal{B}(B\to K \nu \bar \nu)}{ \mathcal{B}(B\to K \nu \bar \nu)_{\rm SM}}=\frac{ \mathcal{B}(B\to K^* \nu \bar \nu)} 
 {\mathcal{B}(B\to K^* \nu \bar \nu)_{\rm SM}}=
 \frac{\mathcal{B}(B\to X_s \nu \bar \nu)}{\mathcal{B}(B\to X_s \nu \bar \nu)_{\rm SM}}=\frac{ |X_{\rm L}(B_s)|^2}{|\eta_X X_0(x_t)|^2}~, 
 \ee
with $X_{\rm L}(B_s)$ given in (\ref{XLB}). The equality of these three ratios 
is an important test of the LHS scenario. The violation of them would 
imply the presence of right-handed couplings at work 
\cite{Colangelo:1996ay,Buchalla:2000sk,Altmannshofer:2009ma}. In the context 
of $Z^\prime$ models this is clearly seen in Fig.~20 of \cite{Buras:2012jb}.

Using the $SU(2)_L$ relation in (\ref{nunu910}) we can now correlate these 
ratios with $\overline{\mathcal{B}}(B_s\to\mu^+\mu^-)$ and $C_9^{\rm NP}$. 
The power of this relation allows us to avoid the discussion of muon couplings 
at present even if knowing them would allow to correlate  $C_9^{\rm NP}$ and 
 $C_{10}^{\rm NP}$.

In Fig.~\ref{fig:BsmuRnu} we show $R_{\nu\bar\nu}$ versus $\overline{\mathcal{B}}(B_s\to\mu^+\mu^-)$ for  $C_9^\text{NP} = -1.6$ (magenta)
and
$C_9^\text{NP} = -0.8$ (cyan) and $-0.8\leq C_{10}^\text{NP}\leq 1.8$. We observe  an anti-correlation between these branching ratios as
opposed to
correlation found in \cite{Buras:2012jb}. But in the latter paper
$\Delta_{L}^{\nu\bar\nu}(Z')=\Delta_A^{\mu\bar\mu}(Z')=0.5$ has been assumed.
With the values of  $C_9^\text{NP}$ as used here these two couplings have
opposite sign and anti-correlation follows. But the predicted NP effects in
$R_{\nu\bar\nu}$ are rather small. The same conclusion has been reached in
\cite{Altmannshofer:2013foa}.

\subsection{Comments on other NP models}
The decays $B_{s,d}\to\mu^+\mu^-$ have been studied in various models in the 
literature but in view of many parameters involved very often no clear cut 
conclusions can be made. Here we just want to mention  four models where this 
can be done. 

In the Littlest Higgs model with T-parity,  $\overline{\mathcal{B}}(B_{s}\to\mu^+\mu^-)$ can only be enhanced with respect to its SM value and this enhancement comes dominantly from the T-even sector \cite{Blanke:2009am}. Larger effects are possible in the 
case of $B_d\to \mu^+\mu^-$ and the CMFV relation (\ref{CMFV6}) as seen in Fig.~8 of 
 \cite{Blanke:2009am} can be significantly violated, in particular for SM-like values of $S_{\psi\phi}$. Thus in LHT
\be
R_{\mu\mu}(B_s)\equiv\frac{\mathcal{B}(B_{s}\to\mu^+\mu^-)}{\mathcal{B}(B_{s}\to\mu^+\mu^-)_{\rm SM}}=
1.15\pm 0.10, \qquad 0.5\le r \le 1.1
\ee
showing that in this model $r< 1$ is favoured over $r>1$ as indicated by 
the CMS and LHCb result in (\ref{rexp}) although the values of $r$ as given 
there cannot be reached. On the other hand the predicted enhancement of 
 $\overline{\mathcal{B}}(B_{s}\to\mu^+\mu^-)$ could turn out to be a problem 
for this model if data improve. We should remark that the operator structure 
in this model is as in LHS and it is a non-MFV model. But tree-level FCNCs 
are absent in this model.

Even if the presence of the fourth generation is unlikely or even excluded 
 \cite{Eberhardt:2012gv} it is interesting to observe that in the case of $S_{\psi\phi}$ 
as found by the LHCb  $\overline{\mathcal{B}}(B_{s}\to\mu^+\mu^-)$ is most 
likely suppressed and  $\mathcal{B}(B_{d}\to\mu^+\mu^-)$ enhanced so that 
the present data on these two branching ratios can be reproduced 
in this model. This is clearly seen in Figs.~4 and 5 of \cite{Buras:2010pi}. 
Thus if not for the difficulties of this model discussed in  \cite{Eberhardt:2012gv} 
the recent data on these decays could be a support for this model. For 
a recent discussion in this spirit see \cite{Hou:2013kia}.

 Next in the Randall-Sundrum model with custodial
protection one finds~\cite{Blanke:2008yr}
\be
R_{\mu\mu}(B_s)=1.00\pm 0.10, \qquad 0.6\le r \le 1.35
\ee
with equal probability of $r$ being suppressed or enhanced with respect to $r=1$.  Finally similar size of
 departures from CMFV have been identified recently 
within 331 models as can be seen in Fig.~12 of \cite{Buras:2013dea}.

\subsection{Comments on the size of reduced couplings}
Our analysis did not make any assumptions on the diagonal couplings of 
$Z^\prime$ to quarks and in the case of charged leptons we did not assume the universality of lepton couplings so that $Z^\prime$ couplings to muons and electrons could be in principle different from each other. In fact as our recent analysis \cite{Buras:2013dea} shows, this violation of lepton universality is required for most interesting cases considered here by us as otherwise only 
the  values $\bar\Delta_A^{\mu\bar\mu}(Z')\le 0.55$ would be allowed
by LEP-II data \cite{Schael:2013ita}. For the vector couplings this bound is even stronger $\bar\Delta_V^{\mu\bar\mu}(Z')\le 0.35$. We refer to section 7.2 of  
\cite{Buras:2013dea} for more details. These findings imply that the large enhancement of $\mathcal{B}(B_{d}\to\mu^+\mu^-)$ in LHS modes 
must be accompanied 
with breakdown of universality in $Z^\prime$ couplings to leptons.

\section{The case of the SM $Z$}\label{sec:4a}
We will next consider the case of the SM $Z$ boson with flavour violating 
couplings. An extensive analysis of this case has been performed in \cite{Buras:2012jb} and it is of interest to see how this scenario faces new data. 
In this case we have
\be
M_Z=91.2\gev, \quad \Delta_L^{\nu\bar\nu}(Z)=\Delta_A^{\mu\bar\mu}(Z)=0.372,
\quad  \Delta_V^{\mu\bar\mu}(Z)=-0.028
\ee
and consequently the reduced leptonic couplings are fixed:
\be\label{Zreduced}
\bar\Delta_L^{\nu\bar\nu}(Z)=\bar\Delta_A^{\mu\bar\mu}(Z)=4.04/\tev,
\quad  \bar\Delta_V^{\mu\bar\mu}(Z)=-0.304/\tev.
\ee

\begin{figure}[!tb]
 \centering
\includegraphics[width = 0.45\textwidth]{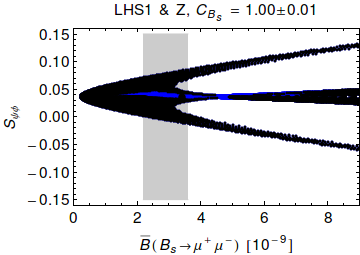}
\includegraphics[width = 0.45\textwidth]{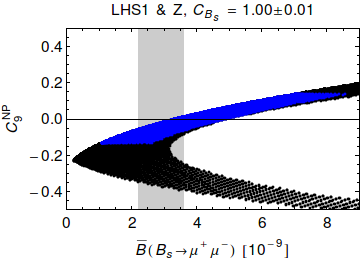}
\caption{ $S_{\psi\phi}$ versus $\overline{\mathcal{B}}(B_{s}\to\mu^+\mu^-)$ (left) and $C_9^\text{NP}$ versus
$\overline{\mathcal{B}}(B_{s}\to\mu^+\mu^-)$ (right) for $C_{B_s} = 1.00\pm0.01$. The black points violates the
bounds 
for $C_{10}^\text{NP}$ (see Eq.~(\ref{equ:C10NPconstraints})).
}\label{fig:Zbin5plot}~\\[-2mm]\hrule
\end{figure}

We observe that $\bar\Delta_L^{\nu\bar\nu}(Z)=\bar\Delta_A^{\mu\bar\mu}(Z)$ is 
much larger than considered presently by us in the case of $Z^\prime$, while as we will soon see 
$\bar\Delta_V^{\mu\bar\mu}(Z)$
turns out to be too small to give the values $C_9^{\rm NP}$ in (\ref{ANOM}).
 In fact as seen in the right panel of Fig.~\ref{fig:Zbin5plot} its most 
negative  value is around $-0.14$.  It is this value that we have used 
in Fig.~\ref{fig:BsmuRnu} to show that in the $Z$-case the effects in 
$b\to s\nu\bar\nu$ transitions are even smaller than in the $Z^\prime$-case.

What remains to be done is to fix the FCNC couplings of $Z$ to quarks by 
imposing the constraints from $\Delta M_s$ and $S_{\psi\phi}$.
This has been already done in Section~9 in \cite{Buras:2012jb} with the 
result that $\overline{\mathcal{B}}(B_s\to\mu^+\mu^-)$ is always larger than
its SM value and mostly above the data known at that time that decreased 
significantly since then. Moreover it has been shown that the constraints 
(\ref{equ:C10NPconstraints}) could not be satisfied.
 However, in \cite{Buras:2012jb} $C_{B_s}=0.927$ 
has been used as this was hinted by lattice data at that time. Requiring  
the agreement with the data on $\Delta M_s$ within $\pm5\%$ implied 
the $Zbs$ coupling to be too large in the presence of a large coupling
$\bar\Delta_A^{\mu\bar\mu}(Z)$ in (\ref{Zreduced}) so that these constraints could not be satisfied. 
But with the new lattice input even  $C_{B_s}=1.00\pm 0.01$ is fine 
and the coupling $\bar\Delta_L^{sb}(Z)$ is allowed to be much smaller so that 
data on $\Delta F=2$ observables and  $\overline{\mathcal{B}}(B_s\to\mu^+\mu^-)$  can be satisfied while being consistent with the 
constraints in (\ref{equ:C10NPconstraints}).  However it turns out 
that only the case of  $C_{B_s}=1.00\pm 0.01$ is admitted and this implies 
that not only $\Delta M_s$ but also $S_{\psi\phi}$ has to be close to their 
SM values (see Fig.~\ref{fig:Zbin5plot}). Still as we will see  significant 
deviations of $\overline{\mathcal{B}}(B_s\to\mu^+\mu^-)$ from the SM prediction are possible because of very large value of $\bar\Delta_A^{\mu\bar\mu}(Z)$.
Improved lattice calculations will tell us whether this 
scenario  works.

\begin{figure}[!tb]
 \centering
\includegraphics[width = 0.45\textwidth]{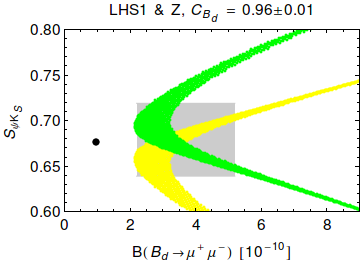}
\includegraphics[width = 0.45\textwidth]{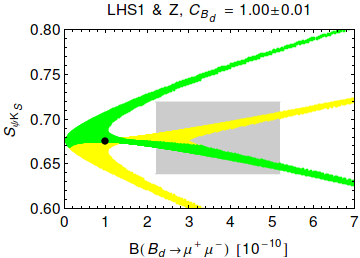}

\includegraphics[width = 0.45\textwidth]{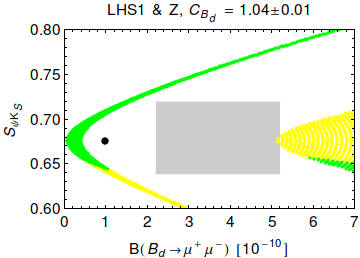}
\caption{ $S_{\psi K_S}$ versus ${\mathcal{B}}(B_{d}\to\mu^+\mu^-)$ for $C_{B_s} = 0.96\pm0.01$, $C_{B_s} = 1.00\pm0.01$ and $C_{B_s} =
1.04\pm0.01$ in LHS1. The yellow and green points correspond to the different oases that differ by $\pi$ in $\delta_{13}$. 
}\label{fig:ZbinplotBdlowVub}~\\[-2mm]\hrule
\end{figure}

\begin{figure}[!tb]
 \centering
\includegraphics[width = 0.45\textwidth]{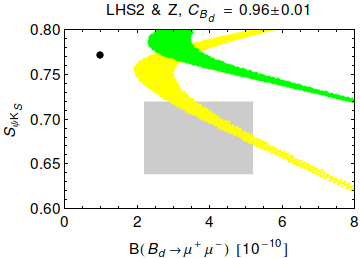}
\includegraphics[width = 0.45\textwidth]{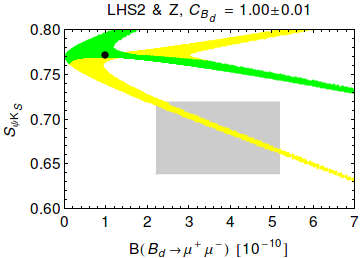}

\includegraphics[width = 0.45\textwidth]{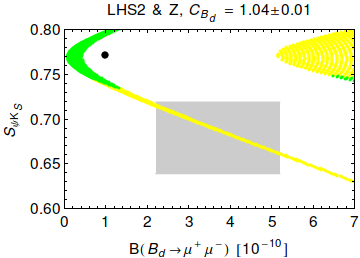}
\caption{ $S_{\psi K_S}$ versus ${\mathcal{B}}(B_{d}\to\mu^+\mu^-)$ for $C_{B_d} = 0.96\pm0.01$, $C_{B_d} = 1.00\pm0.01$ and $C_{B_d} =
1.04\pm0.01$ in LHS2. The yellow and green points correspond to the different oases that differ by $\pi$ in $\delta_{13}$. 
}\label{fig:ZbinplotBdhighVub}~\\[-2mm]\hrule
\end{figure}
Let us then study the case of 
the $B_d\to\mu^+\mu^-$ decay. As now 
the coupling $\bar\Delta_A^{\mu\bar\mu}(Z)$ is fixed it is easier to 
see what happens for different values of $C_{B_d}$. In \cite{Buras:2012jb} 
also $C_{B_d}=0.927$  has been considered implying large enhancements of 
 ${\mathcal{B}}(B_{d}\to\mu^+\mu^-)$, which as seen in Fig.~28 of that paper are  in the LHS1 scenario on top of the CMS and LHCb data but a bit higher, 
although still consistent with the latter, for the LHS2. We could even 
claim that our prediction has been confirmed by  CMS and  LHCb 
data but in view of  large experimental errors and modified lattice results we will update our analysis.

As we fixed $C_{B_s}=1.00\pm 0.01$, $C_{B_d}$ should not differ by too much 
from this value in order to agree with the data on the 
ratio $\Delta M_s/\Delta M_d$.
Therefore we consider only the cases
\be
 C_{B_d}=0.96\pm0.01, \qquad C_{B_d}=1.00\pm 0.01, \qquad C_{B_d}=1.04\pm 0.01~.
\ee
and show for them in Figs.~\ref{fig:ZbinplotBdlowVub} and~\ref{fig:ZbinplotBdhighVub} the correlation between $S_{\psi K_S}$ and
${\mathcal{B}}(B_{d}\to\mu^+\mu^-)$ 
in LHS1 and LHS2 scenarios for $\vub$.  Here we also distinguish between the two different oases (yellow and green points).  The
following observations can be 
made on the basis of these results:
\begin{itemize}
\item
In LHS1 for $C_{B_d}=0.96\pm0.01$  and $C_{B_d}=1.00\pm0.01$ a very good agreement with experiment for both oases can be obtained, but only for  $C_{B_d}=0.96\pm0.01$ is ${\mathcal{B}}(B_{d}\to\mu^+\mu^-)$ forced to be enhanced. For  $C_{B_d}=1.04\pm0.01$ 
it is outside the gray area.
\item
In LHS2 where NP must be present to reduce the value of $S_{\psi K_S}$, 
there is an agreement  for all three values of $C_{B_d}$ but only in the 
yellow oasis (low $\delta_{13}$). Note that in the $Z$-case the muon 
coupling is fixed and  it is  the yellow oasis which is chosen by the data.
\end{itemize}

\begin{figure}[!tb]
 \centering
\includegraphics[width = 0.45\textwidth]{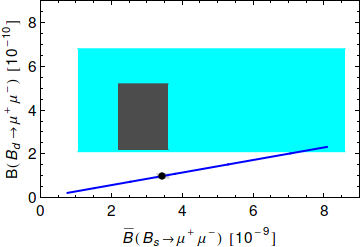}
\includegraphics[width = 0.45\textwidth]{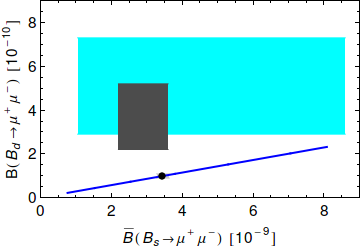}
\caption{ ${\mathcal{B}}(B_{d}\to\mu^+\mu^-)$ versus ${\mathcal{\bar{B}}}(B_{s}\to\mu^+\mu^-)$ for $\vub = 0.0034$ (left) and $\vub =
0.0040$ (right) and  $C_{B_d} = 0.96\pm 0.01$, $C_{B_s} = 1.00\pm 0.01$, $0.639\leq
S_{\psi K_s}\leq 0.719$ and $-0.15\leq S_{\psi\phi}\leq 0.15$. SM is represented by the light gray area with black dot. Dark gray
 region: Combined exp 1$\sigma$
 range
 $\overline{\mathcal{B}}(B_s\to\mu^+\mu^-) = (2.9\pm0.7)\cdot 10^{-9}$ and $\mathcal{B}(B_d\to\mu^+\mu^-) = (3.6^{+1.6}_{-1.4})\cdot
10^{-10}$.}\label{fig:ZBdvsBsLHS}~\\[-2mm]\hrule
\end{figure}

We conclude therefore that the $Z$ boson with left-handed flavour violating couplings can reproduce the CMS and LHCb data for $B_{s,d}\to\mu^+\mu^-$ while being consistent with present
constraints from $\Delta M_d$ and $S_{\psi K_S}$. Interestingly in the LHS1 
scenario there is a preference for $C_{B_d}\le 1.0$ while all values work in the case of LHS2. If the branching ratio decreases in the future below 
$2\cdot 10^{-10}$, LHS1 will be favoured but in that case the lower limit 
on $C_{B_d}$ will be higher than $0.96$.

The result analogous to the plots in  Fig.~\ref{fig:BdvsBsLHS} is shown in 
Fig.~\ref{fig:ZBdvsBsLHS}. As an example we have chosen  $C_{B_d}=0.96\pm0.01$ and  $C_{B_s}=1.00\pm0.01$ for both LHS1 and LHS2. For LHS1
and LHS2 we find respectively
\be 
2.1\cdot 10^{-10}\le\mathcal{B}(B_{d}\to\mu^+\mu^-)\le 6.8\cdot 10^{-10}, \qquad 
{(\rm LHS1)}.
\ee
\be 
2.9\cdot 10^{-10}\le\mathcal{B}(B_{d}\to\mu^+\mu^-)\le 7.3\cdot 10^{-10}, \qquad
{(\rm LHS2)}.
\ee
and in both cases
\be 
1.2\cdot 10^{-10}\le\mathcal{B}(B_s\to\mu^+\mu^-)\le 7.2\cdot 10^{-10}.
\ee

Compared with the plots  in Fig.~\ref{fig:BdvsBsLHS},  
these plots could imply that it is more natural to reproduce the present 
data for $B_{d}\to\mu^+\mu^-$ in the case of $Z$ rather than $Z^\prime$ 
scenario. However we should remember that as seen in Figs.~\ref{fig:binBdlowVub} and 
\ref{fig:binBdhighVub} higher values of $B_{d}\to\mu^+\mu^-$ branching ratios 
can be obtained in the latter scenario by increasing the coupling $\bar\Delta_A^{\mu\bar\mu}(Z^\prime)$ above the unity. 

So far so good. In view of the 
condition  $C_{B_s}=1.00\pm 0.01$ required by the constraints~(\ref{equ:C10NPconstraints}) combined with  the $B_s\to\mu^+\mu^-$ data 
the couplings $\bar\Delta_L^{sb}(Z)$ are too small to compensate the 
smallness of $\bar\Delta_V^{\mu\bar\mu}(Z)$ in the evaluation of $C_{9}^{\rm NP}$ (see Fig.~\ref{fig:Zbin5plot}).
Consequently, the values in (\ref{ANOM}) cannot be reproduced.
Thus the only hope for the $Z$-scenario is a very significant reduction 
of $C_{9}^{\rm NP}$ in the future.

\boldmath
\section{Comments on a real  $C_9^{\rm NP}$ and right-handed couplings}\label{sec:5}
\unboldmath
\boldmath
\subsection{Real $C_9^{\rm NP}$}
\begin{figure}[!tb]
 \centering
\includegraphics[width = 0.45\textwidth]{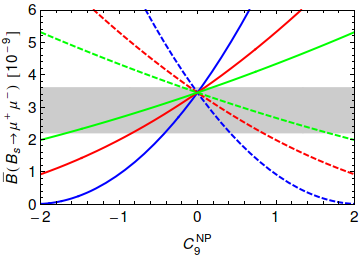}
\includegraphics[width = 0.48\textwidth]{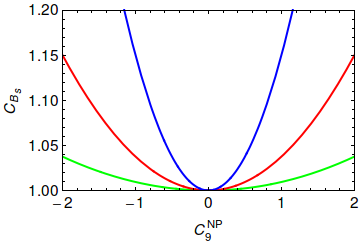}
\caption{  ${\mathcal{\bar{B}}}(B_{s}\to\mu^+\mu^-)$  and $C_{B_s}$ versus a real $C_9^\text{NP}$. Left:
$\bar\Delta_A^{\mu\bar\mu}/\bar\Delta_V^{\mu\bar\mu} = -2$ (blue dashed), $-1$ (red dashed), $-0.5$ (green dashed), 0.5 (green), 1 (red) and
2 (blue). Right: $\bar\Delta_V^{\mu\bar\mu} = \pm0.5~\text{TeV}^{-1}$ (blue), $\pm1$~TeV$^{-1}$ (red) and $\pm2$~TeV$^{-1}$
(green).}\label{fig:C9real}~\\[-2mm]\hrule
\end{figure}

\unboldmath
The recent $B\to K^*\mu^+\mu^-$ anomalies imply according to \cite{Descotes-Genon:2013wba,Altmannshofer:2013foa} the range for  $C_9^{\rm NP}(B_s)$
given in (\ref{ANOM})
and moreover this contribution could come from a $Z^\prime$ exchange. Here we would like to collect the implications of this possibility 
for the LHS. These are
\begin{itemize}
\item 
Unique enhancement of $\Delta M_s$ with respect to its SM value 
implying that this scenario can only be valid for $C_{B_s}> 1.0$.  As we 
have seen in the previous section this is not favoured by the present data 
on $B_{s,d}\to\mu^+\mu^-$ but cannot be excluded due to large errors on 
experimental  $B_{s,d}\to\mu^+\mu^-$ branching ratios. Future 
lattice calculations will tell us whether  $C_{B_s}> 1.0$ is true.  In fact 
the most recent values in (\ref{UTfit}) favour slightly such values.
\item
As in this case $\delta_{23}=\beta_s $  or $\delta_{23}=\beta_s+\pi $, the asymmetry $S_{\psi\phi}$ equals the 
SM one. We are in the  scenario  for SM-like $S_{\psi\phi}$ 
with the restriction 
 $C_{B_s}> 1.0$ and these are the {\it magenta} points
in Figs.~\ref{fig:BsDeltaF2} and~\ref{fig:Bstan}. 
Consequently the CP-asymmetries 
$A_7$ and $A_8$ in $B\to K^*\mu^+\mu^-$ vanish. 
\item
Due to the relation (\ref{AV}) there is a strict correlation between 
$\overline{\mathcal{B}}(B_{s}\to\mu^+\mu^-)$  and $C_9^{\rm NP}(B_s)$ that 
depends on the values of the ratio  $\bar\Delta_A^{\mu\bar\mu}(Z')/\bar\Delta_V^{\mu\bar\mu}(Z')$. We show this correlation in the left panel of 
 Fig.~\ref{fig:C9real}. In the right panel we show using~(\ref{SC9}) $C_{B_s}$
 as a function of a real $C_9^{\rm NP}(B_s)$ for different
values of
$\bar\Delta_V^{\mu\bar\mu}(Z')$ 
 so that some correlation between 
$\Delta M_s$ and $\overline{\mathcal{B}}(B_{s}\to\mu^+\mu^-)$ is present. 
The main message from this plot is that combined data for $\overline{\mathcal{B}}(B_{s}\to\mu^+\mu^-)$  and $C_9^{\rm NP}$ favour
\be
0\le\frac{\bar\Delta_A^{\mu\bar\mu}(Z')}{\bar\Delta_V^{\mu\bar\mu}(Z')}\le 1.0,
\ee
implying that these two couplings should have the same sign.
\item
While in the $B_s$ system there are some similarities of this scenario with the CMFV models, 
LHS differs in the presence of a real  $C_9^{\rm NP}(B_s)$ from CMFV as 
NP physics with new complex phases can enter $B_d$ and $K$ systems. Moreover 
as we have seen the present data on $\mathcal{B}(B_{d}\to\mu^+\mu^-)$ 
favour $\bar\Delta_A^{\mu\bar\mu}(Z')\approx 1.0$ and this can also be correlated  with the results in Fig.~\ref{fig:C9real}.
\end{itemize}

\subsection{Right-handed currents}
In \cite{Buras:2012jb} we have analyzed in addition to LHS scenario also 
RHS scenario in which only right-handed $Z^\prime$ couplings where present 
and two scenarios (LRS and ALRS) with both left-handed and right-handed couplings satisfying the relations
\be
\Delta_L^{qb}(Z^\prime)=\Delta_R^{qb}(Z^\prime), \qquad \Delta_L^{qb}(Z^\prime)=-\Delta_R^{qb}(Z^\prime),
\ee
 respectively.
Our analysis included complex couplings but in the spirit of this section and 
to relate to the analyses in \cite{Descotes-Genon:2013wba,Altmannshofer:2013foa} let us assume here that these couplings have the same phases modulo $\pi$ 
so that resulting Wilson coefficients remain real.

To an excellent approximation the shift $\Delta S$ in (\ref{Zprime1}) is now 
obtained by making the following replacement
\be\label{M12Znew}
(\Delta_L^{bs}(Z^\prime))^2\rightarrow (\Delta_L^{bs}(Z^\prime))^2+ (\Delta_R^{bs}(Z^\prime))^2 +\kappa\Delta_L^{bs}(Z^\prime)\Delta_R^{bs}(Z^\prime),
\ee
where  
\be\label{kappa}
\kappa=2 
\frac{\langle Q_1^\text{LR}(\mu_{Z^\prime})\rangle}{\langle Q_1^\text{VLL}(\mu_{Z^\prime})\rangle} \approx -10.3
\ee
exhibits the known fact that the matrix elements of LR operators are 
strongly enhanced with respect to VLL operators that are solely responsible 
for $\Delta F=2$ effects in the SM and LHS scenario. We would like to add 
that due to different anomalous dimensions of LR and VLL operators, $\kappa$ 
increases with increasing $\mu_{Z^\prime}=\ord(M_{Z^\prime})$. The value given 
in (\ref{kappa}) corresponds to $\mu_{Z^\prime}=1\tev$. For more details, in 
particular NLO corrections, we refer to  \cite{Buras:2013ooa,Buras:2012fs}.
Analogous expressions for other meson systems exist. Now as seen in 
Table~2 of  \cite{Buras:2013ooa} model independently $\kappa$ is {\it negative}
which has an impact on the signs and size of 
NP contributions to $\Delta M_s$.

We make the following observations following more detailed discussion in 
 \cite{Buras:2013ooa,Buras:2012jb}:
\begin{itemize}
\item
In the case of the RHS scenario the $\Delta F=2$ constraints are exactly the same 
as in the LHS scenario but this time on the right-handed couplings.
\item
In the case of the LRS scenario   with the NP phase aligned to the SM one, NP contribution to 
$\Delta M_s$ is strictly {\it negative} as opposed to the LHS scenario 
and because of the $\kappa$-enhancement the coupling $\bar\Delta_L^{sb}(Z^\prime)$
must be suppressed by a factor $\sqrt{8}$ relatively to the LHS or RHS case.
 For fixed leptonic couplings NP effects in  $b\to s \mu^+\mu^-$ transitions are expected to be smaller than in LHS scenario.
As the contribution is negative anyway, in the case of 
establishing $C_{B_s}> 1.00$ this scenario will be ruled out. 
\item 
In the case of ALRS scenario    with the NP phase aligned to the SM one,
NP contribution to 
$\Delta M_s$ is strictly {\it positive} as in  the LHS scenario but
because of the $\kappa$-enhancement the coupling $\bar\Delta_L^{sb}(Z^\prime)$
must be suppressed by a factor $\sqrt{12}$ relatively to the LHS or RHS case.
 For fixed leptonic couplings NP effects in  $b\to s \mu^+\mu^-$ transitions are expected to be smaller than in LHS scenario.
As the contribution is positive  in the case of 
establishing $C_{B_s}<1.00$ this scenario will be ruled out. 
\end{itemize}

Concerning $B_{s,d}\to\mu^+\mu^-$, as emphasized in \cite{Buras:2012jb}, NP 
effects vanish in the LRS scenario independently of the leptonic couplings.
But it could be that the quark couplings in $B_{s}\to\mu^+\mu^-$ 
belong to the LRS scenario and this is the reason why the data are close 
to the SM prediction, while this is not the case for the quark couplings in 
$B_{d}\to\mu^+\mu^-$. In the case of ALRS scenario NP does 
contribute to $B_{s,d}\to\mu^+\mu^-$ and the fact that LH and RH couplings 
enter the decay amplitude with the same sign compensates partly the 
suppression of quark  couplings due to enhanced $\Delta F=2$ matrix elements.

Let us next investigate 
$B_d\to K^*\mu^+\mu^-$ in these scenarios.  We concentrate here 
on three angular observables $F_L$, $S_4$ and $S_5$ introduced in \cite{Altmannshofer:2008dz} that are particularly sensitive to NP
contributions. Useful approximate formulae for them  in terms of  Wilson coefficients have been recently 
 presented in \cite{Altmannshofer:2013foa}. These formulae neglect 
 interferences between NP contributions which is justified in view of 
small room left for these contributions in the data. Applied 
to $Z^\prime$ models these formulae are as follows:
\begin{align}
&\langle F_L \rangle_{[1,6]} \approx 0.77 +0.05 C_9^{\rm NP}-0.04 C_9^\prime+0.04 C_{10}^\prime \label{AS1}\\
&\langle S_4 \rangle_{[14.2,16]} \approx 0.29 -0.02 C_9^\prime+0.03 C_{10}^\prime.
\label{AS2}  \\
& \langle S_5 \rangle_{[1,6]} \approx -0.14 -0.09 C_9^{\rm NP}-0.03 C_9^\prime+0.10 C_{10}^\prime. \label{AS3}
\end{align}
 Here the subscripts on the l.h.s of these formulae indicate for which bin in 
$q^2$ these equations and corresponding data given below apply. In order not to 
obuse notations we will drop these subscripts in what follows.

Also in RHS, LRS and ALRS scenarios, NP contributions to the dipole operator can be neglected \cite{Buras:2012jb}   and this reduction of the number of relevant Wilson
coefficients 
together with the absence of new CP-violating phases allows to derive 
correlations  between these three observables in question 
as we will see soon.
Here the primed coefficients are obtained from the unprimed ones by replacing
$\Delta_L^{qb}(Z^\prime)$ by $\Delta_R^{qb}(Z^\prime)$. 

It should be emphasized that the numerical values in (\ref{AS1})-(\ref{AS3}) 
are subject
to hadronic uncertainties. Therefore, even if in our  numerical analysis of 
various correlations we will use these values, we collect 
in the Appendix~\ref{app:corr}  general
formulae for these correlations. This should allow to update these correlations 
if the numerical values in  (\ref{AS1})-(\ref{AS3}) will be modified. Moreover, 
from the Appendix~\ref{app:corr} one can derive correlations for the 
basis of observables proposed in \cite{Descotes-Genon:2013vna} by using the dictionary \cite{Altmannshofer:2013foa}
\be
S_3=\frac{1}{2} F_T P_1,\qquad S_4=\frac{1}{2} F_{LT} P_4^\prime, \qquad 
S_5=\frac{1}{2} F_{LT} P_5^\prime,
\ee
where $F_{LT}=\sqrt{F_L(1-F_L)}$.

The first terms in    (\ref{AS1})-(\ref{AS3})  are SM 
predictions. The estimate of uncertainties vary from paper to paper. We quote 
here the results from \cite{Altmannshofer:2013foa} 
\be\label{equ:SMvalues}
\langle F_L \rangle= 0.77\pm0.04, \qquad \langle S_4 \rangle=0.29\pm0.07, 
\qquad \langle S_5 \rangle =-0.14\pm0.02.
\ee
For other estimates see \cite{Descotes-Genon:2013vna,Jager:2012uw}. In particular in \cite{Jager:2012uw}  much larger error has been assigned to
$\langle S_5 \rangle$.

Concerning the data,  the ones quoted in  \cite{Altmannshofer:2013foa} 
and given by 
\be\label{equ:expFLS5}
\langle F_L \rangle =0.59\pm0.08, \qquad \langle S_4 \rangle = -0.07\pm0.11,
\qquad \langle S_5 \rangle = 0.10\pm0.10,
\ee
are based on the LHCb data for $P_4^\prime$ and $P_5^\prime$  \cite{Aaij:2013qta} and the average of the data from Belle, Babar, CDF, CMS, ATLAS and LHCb 
on $F_L$ which as discussed in the appendix of \cite{Altmannshofer:2013foa} 
are not in full agreement with each other. We find that 
the  weighted average of the most accurate data from  LHCb \cite{Aaij:2013iag} and CMS \cite{Chatrchyan:2013cda} is $\langle F_L \rangle =0.66\pm0.07$, without basically 
no change in $S_4$ and $S_5$. In our plots we will show both ranges on 
$\langle F_L \rangle$. 
Note that the definition of $S_4$ differs from the LHCb definition by sign.

 The central values, in particular for $S_4$ and $S_5$ differ from SM 
predictions, in particular the sign of $\langle S_5 \rangle$ is opposite. 
But the uncertainties both in theory and experiment are sizable. 
Still this pattern of deviations could be a sign  of NP at work.

 Having this formulae at hand it is evident that in LHS, RHS, LRS and 
    ALRS in which some of the coefficients in question are related to each 
 other correlations between these three angular observables are present. In 
the rest of this section we will exhibit these correlations analytically 
and graphically neglecting all theoretical uncertainties which certainly 
are not small. Our goal is modest: we just want to uncover these correlations 
 as this has not been done in the literature, leaving a sophisticated 
numerical analysis for the future when the data stabilize and a consensus 
between theorists on the hadronic uncertainties will be reached.  We should 
also warn the reader that when our results in $Z^\prime$ models differ 
from SM values by much, the neglect of interferences between different 
NP contributions cannot be fully justified but at the semi-quantitative 
level the presented plots should represent what is going on.

We can now specify the formulae in (\ref{AS1})-(\ref{AS2}) to the four scenarios in question:

\subsubsection*{LHS:}

\begin{align}
&\langle F_L \rangle \approx 0.77 +0.05 C_9^{\rm NP},\\
&
\langle S_4 \rangle \approx 0.29 ,\\
&\langle S_5 \rangle \approx -0.14 -0.09 C_9^{\rm NP}.
\end{align}
 Eliminating $C_9^{\rm NP}$ from these expressions in favour of $\langle S_5\rangle$ we 
find 
\be 
\langle F_L \rangle=0.69-0.56 \langle S_5 \rangle,
\ee
which shows analytically the point made
 in \cite{Descotes-Genon:2013wba,Altmannshofer:2013foa} that
NP effects in $F_L$ and $S_5$ are anti-correlated as observed in the data.

\begin{figure}[!tb]
 \centering
\includegraphics[width = 0.43\textwidth]{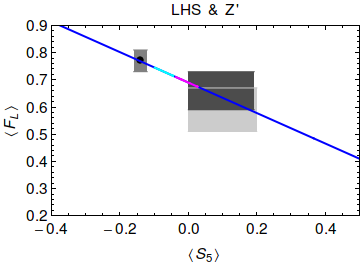}
\includegraphics[width = 0.45\textwidth]{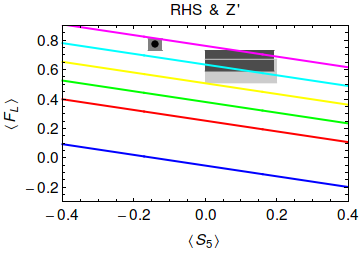}
\includegraphics[width = 0.45\textwidth]{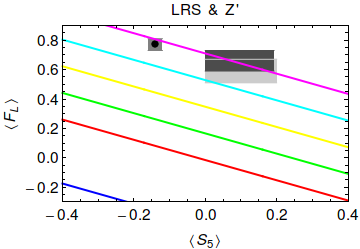}
\includegraphics[width = 0.45\textwidth]{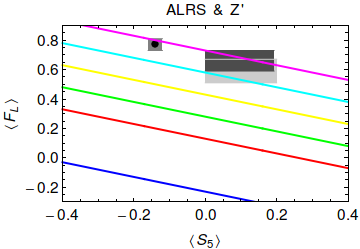}
\caption{$\langle F_L\rangle$ versus $\langle S_5\rangle$ in LHS, RHS, LRS and ALRS. For LHS we have $\langle S_4\rangle = 0$ and the
magenta line corresponds to $C^\text{NP}_9 = -1.6\pm0.3$ and the cyan line to $C^\text{NP}_9 = -0.8\pm0.3$ as
in~(\ref{ANOM}). In the
other scenarios (RHS, LRS, ALRS) we choose $\langle S_4\rangle = -0.02$ (blue, corresponds to the experimental central value), $0.1$ (red),
0.15 (green), 0.2 (yellow), 0.25 (cyan) and 0.3 (magenta). The light gray area corresponds to the experimental range
in~(\ref{equ:expFLS5}) and in the dark gray area we changed $\langle F_L\rangle$ to $0.66\pm0.07$ which also slightly change $\langle
S_{4,5}\rangle$ by a factor $0.963$. The light gray line indicates where the light gray area stops.  The black point and the
gray box correspond to the SM predictions from~(\ref{equ:SMvalues}).}\label{fig:pFLS5LHS}~\\[-2mm]\hrule
\end{figure}

 We show this correlation in the upper left panel  in Fig.~\ref{fig:pFLS5LHS} together with the data. 
We observe that for 
 negative $ C_9^{\rm NP}=\ord(1)$ one can obtain agreement 
with the data for these two observables provided such values are also 
allowed by other constraints. We also observe that for the larger value 
of $\langle F_L\rangle$ (dark grey) it is easier for LHS to describe the 
data. 
On the other hand the value of $\langle S_4 \rangle$ remains SM-like and is significantly larger than indicated by the 
data in (\ref{equ:expFLS5}).  The departure of $\langle S_4 \rangle$ from 
the SM value would be a sign of right-handed currents at work. In
\cite{Altmannshofer:2013foa} it has been concluded that it is difficult 
to find any NP  model 
which could explain the present data on  $\langle S_4 \rangle$ 
and similar conclusion has been reached in \cite{Hambrock:2013zya}. 
However, we will 
keep this observable in our presentation in order to be prepared for future
data.

\subsubsection*{RHS:}
\begin{align}
&\langle F_L \rangle \approx 0.77 -0.04 C_9^\prime+0.04 C_{10}^\prime,\\
&\langle S_4 \rangle \approx 0.29 -0.02 C_9^\prime+0.03 C_{10}^\prime. \\
&\langle S_5 \rangle \approx -0.14 -0.03 C_9^\prime+0.10 C_{10}^\prime.
\end{align}

Eliminating $C^\prime_{10}$ from these expressions we find
\begin{align}
&\langle F_L \rangle=0.826 + 0.40 \langle S_5 \rangle -0.028 C_9^\prime,\\
&\langle F_L \rangle=0.38 + 1.33 \langle S_4 \rangle -0.013 C_9^\prime.
\end{align}

\begin{figure}[!tb]
 \centering
\includegraphics[width = 0.45\textwidth]{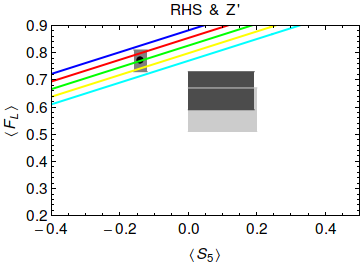}
\includegraphics[width = 0.45\textwidth]{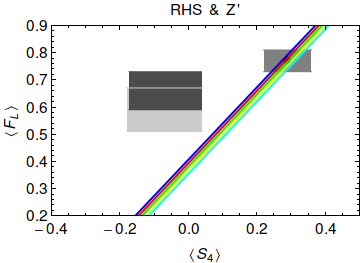}
\caption{$\langle F_L\rangle$ versus $\langle S_5\rangle$ and $\langle S_4\rangle$ in RHS for different values of $C_9^\prime$: $-2$
(blue), $-1$ (red), $0$ (green), $1$ (yellow) and $2$ (cyan). The light gray area corresponds to the experimental range
in~(\ref{equ:expFLS5})  and in the dark gray area we changed $\langle F_L\rangle$ to $0.66\pm0.07$ which also slightly change $\langle
S_{4,5}\rangle$ by a factor $0.963$. The light gray line indicates where the light gray area stops.  The black point and the
gray box correspond to the SM predictions from~(\ref{equ:SMvalues}). }\label{fig:pFLS5RHS}~\\[-2mm]\hrule
\end{figure}

As seen in Fig.~\ref{fig:pFLS5RHS}
NP effects in $F_L$ and $S_5$ are correlated with each other and this scenario 
cannot describe the data even if $ C_9^\prime$ is positive and as large as 2, 
 which is not allowed by other constraints \cite{Altmannshofer:2013foa}. 
Thus in 
agreement with \cite{Descotes-Genon:2013wba,Altmannshofer:2013foa}, right-handed currents alone are not able to explain the anomalies in 
question. 

Finally eliminating $C_9^\prime$ we find a triple correlation 
\be
\langle F_L \rangle=-0.019 -0.36 \langle S_5 \rangle +2.55 \langle S_4 \rangle.
\ee
 We show this correlation in the right upper panel of Fig.~\ref{fig:pFLS5LHS}.
The message in this plot is clear. One cannot reproduce the present data in 
(\ref{equ:expFLS5}).

\subsubsection*{LRS:}
As in this scenario the primed and unprimed coefficients are equal to 
each other we find
\begin{align}
&\langle F_L \rangle \approx 0.77 +0.01 C_9^{\rm NP} +0.04 C_{10}^{\rm NP},\\
&\langle S_4 \rangle \approx 0.29 -0.02 C_9^{\rm NP}+0.03 C_{10}^{\rm NP}.\\
&\langle S_5 \rangle \approx -0.14 -0.12 C_9^{\rm NP}+0.10 C_{10}^{\rm NP},
\end{align}
where we used $C_{10}^{\rm NP}= C_{10}^\prime$. Note, that in spite of the 
appearance of $C_{10}^{\rm NP}$ there is no constraint from $B_s\to\mu^+\mu^-$ 
as NP contribution to this decay 
vanishes in this scenario, except that if future data 
will disagree with the SM prediction for this decay, LRS will not be able 
to explain this.

\begin{figure}[!tb]
 \centering
\includegraphics[width = 0.45\textwidth]{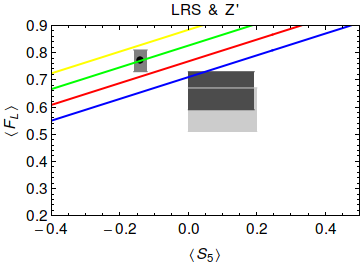}
\includegraphics[width = 0.45\textwidth]{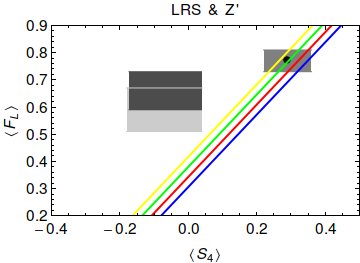}
\caption{$\langle F_L\rangle$ versus $\langle S_5\rangle$ and $\langle S_4\rangle$ in LRS for different values of $C_9^\text{NP}$: $-2$
(blue), $-1$ (red), $0$ (green) and $1$ (yellow). The light gray area corresponds to the experimental range
in~(\ref{equ:expFLS5}) and in the dark gray area we changed $\langle F_L\rangle$ to $0.66\pm0.07$ which also slightly change $\langle
S_{4,5}\rangle$ by a factor $0.963$. The light gray line indicates where the light gray area stops.  The black point and the
gray box correspond to the SM predictions from~(\ref{equ:SMvalues}).}\label{fig:pFLS5LRS}~\\[-2mm]\hrule
\end{figure}

Eliminating $C_{10}^{\rm NP}$ from these expressions we find
\begin{align}
&\langle F_L \rangle=0.826 + 0.40 \langle S_5 \rangle +0.058 C_9^{\rm NP},\\
&\langle F_L \rangle=0.38 + 1.33 \langle S_4 \rangle +0.037 C_9^{\rm NP}.
\end{align}
We show these correlations  in Fig.~\ref{fig:pFLS5LRS}, where different 
colours represent different values of  $C_9^{\rm NP}$. Evidently, it is 
impossible to describe the data in these two plots simultaneously with 
the same value of this coefficient. While in the left-plot a sufficiently 
large negative value of $C_9^{\rm NP}$ would help in explaining the data,
in the right plot a positive value is required. This is also seen in Fig.~\ref{fig:pFLS5LHS}
where we show the triple correlation which one obtains after
eliminating   $C_9^{\rm NP}$  
\be
\langle F_L \rangle=-0.38 - 0.688 \langle S_5 \rangle +3.63 \langle S_4 \rangle.
\ee

\subsubsection*{ALRS:}

\begin{figure}[!tb]
 \centering
\includegraphics[width = 0.45\textwidth]{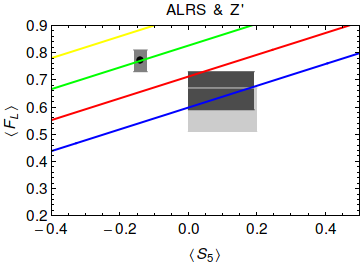}
\includegraphics[width = 0.45\textwidth]{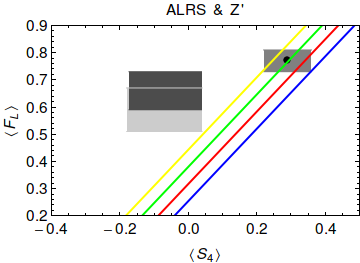}
\caption{$\langle F_L\rangle$ versus $\langle S_5\rangle$ and $\langle S_4\rangle$ in ALRS for different values of $C_9^\text{NP}$: $-2$
(blue), $-1$ (red), $0$ (green) and $1$ (yellow). The light gray area corresponds to the experimental
range
in~(\ref{equ:expFLS5}) and in the dark gray area we changed $\langle F_L\rangle$ to $0.66\pm0.07$. The light gray line indicates where
the light gray area stops.  The black point and the
gray box correspond to the SM predictions from~(\ref{equ:SMvalues}).}\label{fig:pFLS5ALRS}~\\[-2mm]\hrule
\end{figure}

As in this scenario the primed and unprimed coefficients differ by sign
 we find
\begin{align}
&\langle F_L \rangle \approx 0.77 +0.09 C_9^{\rm NP}-0.04  C_{10}^{\rm NP},\\
&\langle S_4 \rangle \approx 0.29 +0.02 C_9^{\rm NP}-0.03 C_{10}^{\rm NP},\\
&\langle S_5 \rangle \approx -0.14 -0.06 C_9^{\rm NP}-0.10  C_{10}^{\rm NP}, 
\end{align}
where we used $C_{10}^{\rm NP}=- C_{10}^\prime$. We find in this case the 
following correlations
\begin{align}
&\langle F_L \rangle=0.826 + 0.40 \langle S_5 \rangle +0.114 C_9^{\rm NP},\\
&\langle F_L \rangle=0.38 + 1.33 \langle S_4 \rangle +0.063 C_9^{\rm NP}.
\end{align}
In fact this scenario is rather close to the one investigated numerically in 
\cite{Altmannshofer:2013foa}.  We show these correlations in Fig.~\ref{fig:pFLS5ALRS}. The one between  $\langle F_L \rangle$ and
$\langle S_4 \rangle$ is 
basically unchanged relatively to the LRS case while the one between 
$\langle F_L \rangle$ and $\langle S_5 \rangle$ appears to be in a better 
shape when compared with the data. In fact as emphasized in \cite{Altmannshofer:2013foa} it is easier to 
obtain the agreement with the data on $\langle F_L \rangle$ and $\langle S_5 \rangle$ by including  a non-vanishing  $C_9^\prime$ in addition to $ C_9^{\rm NP}$. As seen in the left plot 
of Fig.~4 in the latter paper, the data on $A_{\rm FB}$ and $B\to K\mu^+\mu^-$ 
 favour $C_9^\prime$ to be of similar magnitude as $C_9^{\rm NP}$ but having opposite sign and this is our ALRS scenario.

Finally, we find the triple correlation
\be
\langle F_L \rangle=-0.17 - 0.50 \langle S_5 \rangle +3.00 \langle S_4 \rangle,
\ee
which we show in the lower right panel in  Fig.~\ref{fig:pFLS5LHS}. 
Again we observe the problem with describing the data on  $\langle S_4 \rangle$.

 Clearly a complete analysis must include constraints on the values 
of $C_9^{\rm NP}$ from other observables but such an analysis in 
the scenarios with right-handed currents is beyond the scope of the 
present paper. 
We refer to \cite{Altmannshofer:2013foa} for a sophisticated model independent 
numerical 
analysis of such scenarios which also involve the coefficients of the
dipole operators which in $Z^\prime$ models are very suppressed.

The summary of this short excursion in the world of right-handed currents 
is as follows:
\begin{itemize}
\item
The finalists in the competition between these four scenarios are LHS and 
ALRS and in agreement with  \cite{Altmannshofer:2013foa} the present data on
$\langle F_L \rangle$ and $\langle S_5 \rangle$ seem to favour ALRS over LHS.
\item
In order for these finalists to work  the future experimental value 
of $\langle S_4 \rangle$ must be SM-like. This is in particular the case 
of LHS scenario. In ALRS this implies as seen in (\ref{AS2})
\be\label{S4constraint}
 C_{10}^{\rm NP} \approx \frac{2}{3}C_{9}^{\rm NP}
\ee
and the correlation with $B_s\to\mu^+\mu^-$ decay.
We find then that ${\mathcal{\bar{B}}}(B_{s}\to\mu^+\mu^-)\ge 4.6\cdot 10^{-9}$.
\end{itemize}

 In order to understand this result
let us  recall that with real NP contributions to the Wilson coefficients 
a positive (negative) $C_{10}^\prime$ enhances (suppresses) the $B_s\to\mu^+\mu^-$ branching ratio while the opposite is true for $C_{10}^{\rm NP}$. In the 
ALRS scenario  $C_{10}^\prime=-C_{10}^{\rm NP}$ and taking into account that the 
data on $\langle F_L \rangle$ and $\langle S_5 \rangle$ require a negative 
$C_9^{\rm NP}$ implies through (\ref{S4constraint}) an enhancement of 
${\mathcal{\bar{B}}}(B_{s}\to\mu^+\mu^-)$, which could be problematic for 
ALRS one day.

No such correlation is present in LHS but in this scenario as shown at the 
beginning of this section, a real $C_9^{\rm NP}$ implies uniquely an enhancement 
of $\Delta M_s$ and $C_{B_s}> 1.00$, which is not favoured by our analysis of $B_{s,d}\to\mu^+\mu^-$. Solution to this possible 
problem are new CP-violating phases and this could 
be tested in $B_d\to K^*\mu^+\mu^-$ by the measurements of the CP-asymmetries 
$A_7$ and $A_8$.

Finally let us emphasize that the decay $B_d\to K\mu^+\mu^-$ can also 
contribute to this discussion. Indeed the authors of \cite{Altmannshofer:2013foa} provided an approximate formula for the branching ratio confined to 
large $q^2$ region. Lattice calculations of the relevant form factors 
are making significant progress here \cite{Bouchard:2013mia,Bouchard:2013eph} 
and the importance of this decay will increase in the future.
For real Wilson coefficients and neglecting the interference between NP 
contributions the formula of \cite{Altmannshofer:2013foa}  reduces in the absence of NP contributions to 
Wilson coefficients of 
dipole operators to
\be
10^7\times \mathcal{B}(B_d\to K\mu^+\mu^-)_{[14.2,22]}=1.11+ 0.27~(C_9^\text{NP}+C_9^\prime)- 0.27~(C_{10}^\text{NP}+C_{10}^\prime)
\ee
where the error on the first SM term is estimated to be $10\%$ \cite{Bouchard:2013mia,Bouchard:2013eph}. This should 
 be compared with the LHCb result
\be\label{equ:expBdKmu}
10^7\times \mathcal{B}(B_d\to K\mu^+\mu^-)_{[14.2,22]}=1.04\pm 0.12 \qquad {(\rm LHCb).}
\ee

\begin{figure}[!tb]
 \centering
\includegraphics[width = 0.45\textwidth]{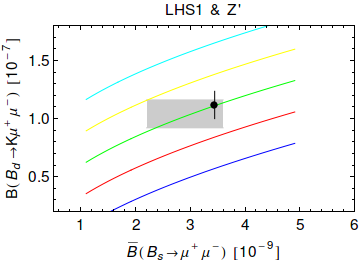}
\caption{$\mathcal{B}(B_d\to K\mu^+\mu^-)$ versus ${\mathcal{\bar{B}}}(B_{s}\to\mu^+\mu^-)$ in LHS for different values of $C_9^\text{NP}$:
$-2$
(blue), $-1$ (red), $0$ (green), $1$ (yellow) and $2$ (cyan) and $-0.8\leq C_{10}^\text{NP}\leq 1.8$. The  gray area corresponds to the
experimental
range in~(\ref{equ:expBdKmu}). SM is represented by the black point.}\label{fig:pBsmuvsBdKmu}~\\[-2mm]\hrule
\end{figure}

We can now explicitly see what happens in the four scenarios discussed by us. 
\begin{itemize}
\item
For LHS we find 
\be\label{LHSK}
10^7\times \mathcal{B}(B_d\to K\mu^+\mu^-)_{[14.2,22]}=1.11+ 
0.27~(C_9^{\rm NP}-C_{10}^{\rm NP}),
\ee
while for RHS the Wilson coefficients $C_{9,10}^{\rm NP}$ should be replaced by
$C_{9,10}^{\prime}$. As RHS is not a favourite scenario we will not consider it 
further.
\item
For LRS we simply have 
\be\label{LRSK}
10^7\times \mathcal{B}(B_d\to K\mu^+\mu^-)_{[14.2,22]}=1.11+ 
0.54~(C_9^{\rm NP}-C_{10}^{\rm NP}).
\ee
In this scenario  NP contributions to $B_s\to\mu^+\mu^-$ vanish but as 
seen in this formula they could  still be visible in $B_d\to K\mu^+\mu^-$ and 
this is a characteristic feature of this scenario.
\item
The opposite takes place in ALRS where NP contributions to $B_s\to\mu^+\mu^-$ 
can be present but vanish for $B_d\to K\mu^+\mu^-$. In 
this scenario there is no constraint on angular observables from 
$B_d\to K\mu^+\mu^-$.
\end{itemize}

As stressed in \cite{Altmannshofer:2013foa} the Wilson coefficient $C_9^{\rm NP}$ by itself has difficulty in removing completely the anomalies in 
$B_d\to K^*\mu^+\mu^-$ due to the constraint from $B_d\to K\mu^+\mu^-$. The LHS 
scenario allows us to have a closer look at this issue. Using (\ref{LHSK}) we show
in Fig.~\ref{fig:pBsmuvsBdKmu}
  the correlation between $\mathcal{B}(B_d\to K\mu^+\mu^-)_{[14.2,22]}$
and ${\mathcal{\bar{B}}}(B_{s}\to\mu^+\mu^-)$ for various values of 
$C_9^{\rm NP}$. We show also the $\pm 10\%$ error on SM prediction which should 
also be taken into account in the lines corresponding to NP predictions with 
$C_9^{\rm NP}\not=0$.
Indeed in agreement with  \cite{Altmannshofer:2013foa}  
only $|C_9^{\rm NP}|\le 1.0$ is allowed at $1\sigma$  which is 
insufficient, as seen in Fig.~\ref{fig:pFLS5LHS}, to remove completely 
$B_d\to K^*\mu^+\mu^-$ anomalies. 

Similarly  our recent analysis \cite{Buras:2013dea} shows that
assuming lepton universality one can derive 
 an upper bound   $|C^{\rm NP}_{9}|\le 1.1 (1.4)$ from
LEP-II data for {\it all}  $Z^\prime$ models with only left-handed flavour
violating couplings to quarks when NP contributions to
$\Delta M_s$ at the level of  $10\%(15\%)$  are allowed. In  concrete dynamical  LHS models, like 331 models analyzed in  \cite{Buras:2013dea}, one even 
finds  $|C^{\rm NP}_{9}|\le 0.8$.

\section{Conclusions and outlook}\label{sec:6}
Motivated by  recent experimental results on $B_{s,d}\to \mu^+\mu^-$ and 
$B\to K^*\mu^+\mu^-$  we have revisited the LHS scenario of \cite{Buras:2012jb}
generalizing it to arbitrary $M_{Z'}$, $\bar\Delta_A^{\mu\bar\mu}(Z')$ and 
departures of $(\Delta M_{s,d})_{\rm SM}$ from the data. This allowed us to present 
new correlations between various observables. These correlations,  shown in many figures  will allow in the coming years to monitor whether
LHS can describe 
future more precise data when also lattice results improve. As already 
emphasized at the beginning of our paper this will certainly be non-trivial.

Beyond these 
correlations possibly the most important results of this paper are the 
following ones:
\begin{itemize}
\item
LHS provides a simple model that allows for the violation of the CMFV relation
between the branching ratios for $B_{d,s}\to \mu^+\mu^-$ and $\Delta M_{s,d}$. 
The plots in Figs.~\ref{fig:BdvsBsLHS} and \ref{fig:ZBdvsBsLHS} for $Z^\prime$ 
and $Z$ illustrate this.
\item
Moreover, LHS is able to accommodate the recent experimental 
results for $B_{s,d}\to \mu^+\mu^-$ but this in the case of 
$\mathcal{B}(B_{d}\to\mu^+\mu^-)$ requires  $\bar\Delta_A^{\mu\bar\mu}(Z')\ge 1.0$ and larger by a factor of two relative to the one
considered by us in \cite{Buras:2012jb}. 
\item
 The SM $Z$ boson with FCNC couplings to quarks can describe
the present data on $B_{s,d}\to\mu^+\mu^-$ provided $\Delta M_s$ and 
$S_{\psi\phi}$ are very 
close to their SM values. However, in view of its small vector coupling to 
muons it cannot describe the anomalies in $B_d\to K^*\mu^+\mu^-$.
\item
The explanation of the present anomalies in $B_d\to K^*\mu^+\mu^-$ with a 
real  $C_9^{\rm NP}(B_s)$ as proposed in \cite{Descotes-Genon:2013wba,Altmannshofer:2013foa}  implies uniquely an enhancement of $\Delta
M_s$ over its SM
value. As the 
present SM value for the ratio $\Delta M_s/\Delta M_d$ agrees with the data 
very well, it is likely that also $\Delta M_d$ should be enhanced in this 
scenario to 
agree with experiment. For this pattern to agree with the data the values of 
the non-perturbative parameters in $\Delta M_{s,d}$ have to be 
lower than their present central values.
\item
We have pointed out that the absence of relevant NP contributions to $C_{7\gamma}$ and $C^\prime_{7\gamma}$ in $Z^\prime$ scenarios
implies in the limit 
of negligible new sources of CP violation correlations  between the 
angular observables $\langle F_L\rangle$, $\langle S_4\rangle$ and 
$\langle S_5\rangle$ in LHS, RHS, LRS and ALRS 
scenarios. In the LHS case there is a unique anti-correlation between 
$\langle F_L\rangle$  and $\langle S_5\rangle$, while in the other cases
the correlations between any of these two observables depend on the value 
of $C_9^{\rm NP}(B_s)$. Eliminating  $C_9^{\rm NP}(B_s)$ in favour of one of these variables results in triple
correlations 
between $\langle F_L\rangle$, $\langle S_4\rangle$ and 
$\langle S_5\rangle$ in RHS, LRS and ALRS scenarios. In LHS NP contributions 
to $\langle S_4\rangle$ can be neglected. These correlations depend on hadronic
uncertainties which should be 
significantly reduced before the correlations in question become really 
useful. Therefore in Appendix~\ref{app:corr} we have presented general 
formulae for correlations which can be efficiently used in the 
flavour precision era.
\item
Our graphical and analytical presentation of correlations between angular 
observables in  $B\to K^*\mu^+\mu^-$ is complementary to the sophisticated 
numerical analyses in  \cite{Descotes-Genon:2013wba,Altmannshofer:2013foa}. 
Among the four $Z^\prime$ scenarios considered by us  ALRS appears to be the favourite scenario followed closely 
by LHS  in agreement with \cite{Altmannshofer:2013foa}. However 
the simple LHS, as advocated in  \cite{Descotes-Genon:2013wba} and in the 
dominant part of the present paper could turn out to be the winner when 
the data and theory will be improved.
\item
Once the Wilson coefficients $C^{\rm NP}_9$ and $C^{\rm NP}_{10}$ will be 
determined through $B_s\to\mu^+\mu^-$, $B\to K^*\mu^+\mu^-$ and $B\to K\mu^+\mu^-$ data the $SU(2)_L$ relation in (\ref{nunu910}) will allow
within the LHS 
model to predict uniquely the $b\to s\nu\bar\nu$ observables. The plot 
in Fig.~\ref{fig:BsmuRnu} illustrates this.
\end{itemize}

Finally, our analysis has shown how important it is to find out what the values 
of $C_{B_s}$ and $C_{B_d}$ from lattice calculations are, measure precisely 
the asymmetries $S_{\psi\phi}$ and $S_{\psi K_S}$ and  improve both theoretical 
and experimental status of all $b\to s \mu^+\mu^-$ observables discussed by us.

We are looking forward to the flavour precision era in which the simple LHS 
scenario which dominated our paper will be much better tested than it is possible now. If it fails, the 
simplest solution would be to introduce right-handed $Z'$ couplings to 
quarks but as reviewed in  \cite{Buras:2013ooa} this  is still another story 
to which we  devoted a short discussion here.

\section*{Acknowledgements}
 We would like to thank Wolfgang Altmannshofer  and David Straub for illuminating comments on 
the status of angular observables and Christoph Bobeth and Cecilia Tarantino for useful informations.
This research was  financed and done in the context of the ERC Advanced Grant project ``FLAVOUR'' (267104) with some support from the DFG cluster of excellence 
''Origin and Structure of the Universe''.

\begin{appendix}
\section{General formulae for correlations}\label{app:corr}
In the limit of no new CP-violating phases and  neglecting the contributions 
of dipole operators as well as interferences between NP contributions one 
can generally write
\begin{align}
&\langle\Delta F_L\rangle\equiv\langle F_L\rangle -\langle F_L\rangle_{\rm SM}=
a_9 C_9^{\rm NP}+a_{9^\prime}C_9^\prime+a_{10^\prime} C_{10}^\prime,\\
& \langle\Delta S_4\rangle\equiv\langle S_4\rangle -\langle S_4\rangle_{\rm SM}=
b_{9^\prime}C_9^\prime+b_{10^\prime} C_{10}^\prime,\\
& \langle\Delta S_5\rangle\equiv\langle S_5\rangle -\langle S_5\rangle_{\rm SM}=
d_9 C_9^{\rm NP}+d_{9^\prime}C_9^\prime+d_{10^\prime} C_{10}^\prime~.
\end{align}
The coefficients $a_i$, $b_i$ and $d_i$ are subject to hadronic uncertainties. 
For the form factors used in \cite{Altmannshofer:2013foa} one has (\ref{AS1})-(\ref{AS3}) but as 
these coefficients will change with time it is useful to present general 
expressions. In particular one
can derive correlations between 
$\langle\Delta F_L\rangle$,
$\langle\Delta S_4\rangle$ and  $\langle\Delta S_5\rangle$. 

In the LHS scenario we simply have
\be
\langle\Delta F_L\rangle=\frac{a_9}{d_9}\langle\Delta S_5\rangle, \qquad 
\langle\Delta S_4\rangle=0~.
\ee

For LR and ALRS scenarios we introduce
\begin{align}
& r_1^\pm=a_9\pm a_{9^\prime} - \frac{a_{10^\prime}}{d_{10^\prime}}(d_9\pm d_{9^\prime})\\
& r_2^\pm=a_9\pm a_{9^\prime} \mp\frac{a_{10^\prime}}{b_{10^\prime}}b_{9^\prime}
\end{align}
with $r_{1,2}^+$ for LRS and $r_{1,2}^-$ for ALRS.

Then we find the correlations
\be
\langle\Delta F_L\rangle=\frac{a_{10^\prime}}{d_{10^\prime}}\langle\Delta S_5\rangle+r_1^\pm C_9^{\rm NP}
\ee
\be
\langle\Delta F_L\rangle=\frac{a_{10^\prime}}{b_{10^\prime}}\langle\Delta S_4\rangle+r_2^\pm C_9^{\rm NP}
\ee
and the triple correlation
\be
\left(1-\frac{r_1^\pm}{r_2^\pm}\right)\langle\Delta F_L\rangle=\frac{a_{10^\prime}}{d_{10^\prime}}\langle\Delta S_5\rangle-
\frac{r_1^\pm}{r_2^\pm} \frac{a_{10^\prime}}{b_{10^\prime}}\langle\Delta S_4\rangle.
\ee
In the RHS scenario one should set $a_9=d_9=0$ and $C_9^{\rm NP}=C_9^\prime$ 
in this formulae and use $r_{1,2}^+$. 

Inserting the numerical values for  $a_i$, $b_i$ and $d_i$ given in (\ref{AS1})-(\ref{AS3}) into these formulae one recovers the
correlations used in our 
numerical analysis.

\end{appendix}

\bibliographystyle{JHEP}
\bibliography{allrefs}

\providecommand{\href}[2]{#2}\begingroup\raggedright\begin{thebibliography}{10}

\bibitem{Buras:2013ooa}
A.~J. Buras and J.~Girrbach, {\it {Towards the Identification of New Physics
  through Quark Flavour Violating Processes}},
  \href{http://xxx.lanl.gov/abs/1306.3775}{{\tt arXiv:1306.3775}}.

\bibitem{Buras:2000dm}
A.~J. Buras, P.~Gambino, M.~Gorbahn, S.~Jager, and L.~Silvestrini, {\it
  Universal unitarity triangle and physics beyond the standard model},  {\em
  Phys. Lett.} {\bf B500} (2001) 161--167,
  [\href{http://xxx.lanl.gov/abs/hep-ph/0007085}{{\tt hep-ph/0007085}}].

\bibitem{Buras:2003jf}
A.~J. Buras, {\it Minimal flavor violation},  {\em Acta Phys. Polon.} {\bf B34}
  (2003) 5615--5668, [\href{http://xxx.lanl.gov/abs/hep-ph/0310208}{{\tt
  hep-ph/0310208}}].

\bibitem{Chivukula:1987py}
R.~S. Chivukula and H.~Georgi, {\it Composite technicolor standard model},
  {\em Phys. Lett.} {\bf B188} (1987) 99.

\bibitem{Hall:1990ac}
L.~J. Hall and L.~Randall, {\it Weak scale effective supersymmetry},  {\em
  Phys. Rev. Lett.} {\bf 65} (1990) 2939--2942.

\bibitem{D'Ambrosio:2002ex}
G.~D'Ambrosio, G.~F. Giudice, G.~Isidori, and A.~Strumia, {\it {Minimal flavour
  violation: An effective field theory approach}},  {\em Nucl. Phys.} {\bf
  B645} (2002) 155--187, [\href{http://xxx.lanl.gov/abs/hep-ph/0207036}{{\tt
  hep-ph/0207036}}].

\bibitem{Kagan:2009bn}
A.~L. Kagan, G.~Perez, T.~Volansky, and J.~Zupan, {\it {General Minimal Flavor
  Violation}},  {\em Phys.Rev.} {\bf D80} (2009) 076002,
  [\href{http://xxx.lanl.gov/abs/0903.1794}{{\tt arXiv:0903.1794}}].

\bibitem{Barbieri:2011ci}
R.~Barbieri, G.~Isidori, J.~Jones-Perez, P.~Lodone, and D.~M. Straub, {\it
  {U(2) and Minimal Flavour Violation in Supersymmetry}},  {\em Eur.Phys.J.}
  {\bf C71} (2011) 1725, [\href{http://xxx.lanl.gov/abs/1105.2296}{{\tt
  arXiv:1105.2296}}].

\bibitem{Barbieri:2012uh}
R.~Barbieri, D.~Buttazzo, F.~Sala, and D.~M. Straub, {\it {Flavour physics from
  an approximate $U(2)^3$ symmetry}},  {\em JHEP} {\bf 1207} (2012) 181,
  [\href{http://xxx.lanl.gov/abs/1203.4218}{{\tt arXiv:1203.4218}}].

\bibitem{Crivellin:2011fb}
A.~Crivellin, L.~Hofer, and U.~Nierste, {\it {The MSSM with a Softly Broken
  $U(2)^3$ Flavor Symmetry}},  {\em PoS} {\bf EPS-HEP2011} (2011) 145,
  [\href{http://xxx.lanl.gov/abs/1111.0246}{{\tt arXiv:1111.0246}}].

\bibitem{Buras:2012sd}
A.~J. Buras and J.~Girrbach, {\it {On the Correlations between Flavour
  Observables in Minimal $U(2)^3$ Models}},  {\em JHEP} {\bf 1301} (2013) 007,
  [\href{http://xxx.lanl.gov/abs/1206.3878}{{\tt arXiv:1206.3878}}].

\bibitem{Altmannshofer:2009ne}
W.~Altmannshofer, A.~J. Buras, S.~Gori, P.~Paradisi, and D.~M. Straub, {\it
  {Anatomy and Phenomenology of FCNC and CPV Effects in SUSY Theories}},  {\em
  Nucl.Phys.} {\bf B830} (2010) 17--94,
  [\href{http://xxx.lanl.gov/abs/0909.1333}{{\tt arXiv:0909.1333}}].

\bibitem{Buras:2012dp}
A.~J. Buras, F.~De~Fazio, J.~Girrbach, and M.~V. Carlucci, {\it {The Anatomy of
  Quark Flavour Observables in 331 Models in the Flavour Precision Era}},  {\em
  JHEP} {\bf 1302} (2013) 023, [\href{http://xxx.lanl.gov/abs/1211.1237}{{\tt
  arXiv:1211.1237}}].

\bibitem{Buras:2012jb}
A.~J. Buras, F.~De~Fazio, and J.~Girrbach, {\it {The Anatomy of Z' and Z with
  Flavour Changing Neutral Currents in the Flavour Precision Era}},  {\em JHEP}
  {\bf 1302} (2013) 116, [\href{http://xxx.lanl.gov/abs/1211.1896}{{\tt
  arXiv:1211.1896}}].

\bibitem{Buras:2013td}
A.~J. Buras, J.~Girrbach, and R.~Ziegler, {\it {Particle-Antiparticle Mixing,
  CP Violation and Rare K and B Decays in a Minimal Theory of Fermion Masses}},
   {\em JHEP} {\bf 1304} (2013) 168,
  [\href{http://xxx.lanl.gov/abs/1301.5498}{{\tt arXiv:1301.5498}}].

\bibitem{Buras:2013uqa}
A.~J. Buras, R.~Fleischer, J.~Girrbach, and R.~Knegjens, {\it {Probing New
  Physics with the $B_s\to\mu^+\mu^-$ Time-Dependent Rate}},  {\em JHEP} {\bf
  1307} (2013) 77, [\href{http://xxx.lanl.gov/abs/1303.3820}{{\tt
  arXiv:1303.3820}}].

\bibitem{Buras:2013rqa}
A.~J. Buras, F.~De~Fazio, J.~Girrbach, R.~Knegjens, and M.~Nagai, {\it {The
  Anatomy of Neutral Scalars with FCNCs in the Flavour Precision Era}},  {\em
  JHEP} {\bf 1306} (2013) 111, [\href{http://xxx.lanl.gov/abs/1303.3723}{{\tt
  arXiv:1303.3723}}].

\bibitem{Langacker:2008yv}
P.~Langacker, {\it {The Physics of Heavy $Z^\prime$ Gauge Bosons}},  {\em
  Rev.Mod.Phys.} {\bf 81} (2009) 1199--1228,
  [\href{http://xxx.lanl.gov/abs/0801.1345}{{\tt arXiv:0801.1345}}].

\bibitem{Barger:2009qs}
V.~Barger, L.~L. Everett, J.~Jiang, P.~Langacker, T.~Liu, {\em et.~al.}, {\it
  {$b\to s$ Transitions in Family-dependent $U(1)^\prime$ Models}},  {\em JHEP}
  {\bf 0912} (2009) 048, [\href{http://xxx.lanl.gov/abs/0906.3745}{{\tt
  arXiv:0906.3745}}].

\bibitem{Fox:2011qd}
P.~J. Fox, J.~Liu, D.~Tucker-Smith, and N.~Weiner, {\it {An Effective Z'}},
  {\em Phys.Rev.} {\bf D84} (2011) 115006,
  [\href{http://xxx.lanl.gov/abs/1104.4127}{{\tt arXiv:1104.4127}}].

\bibitem{Altmannshofer:2011gn}
W.~Altmannshofer, P.~Paradisi, and D.~M. Straub, {\it {Model-Independent
  Constraints on New Physics in $b\to s\gamma$ Transitions}},  {\em JHEP} {\bf
  1204} (2012) 008, [\href{http://xxx.lanl.gov/abs/1111.1257}{{\tt
  arXiv:1111.1257}}].

\bibitem{Altmannshofer:2012ir}
W.~Altmannshofer and D.~M. Straub, {\it {Cornering New Physics in $b\to s$
  Transitions}},  {\em JHEP} {\bf 1208} (2012) 121,
  [\href{http://xxx.lanl.gov/abs/1206.0273}{{\tt arXiv:1206.0273}}].

\bibitem{Dighe:2012df}
A.~Dighe and D.~Ghosh, {\it {How large can the branching ratio of $B_s \to
  \tau^+ \tau^-$ be ?}},  {\em Phys.Rev.} {\bf D86} (2012) 054023,
  [\href{http://xxx.lanl.gov/abs/1207.1324}{{\tt arXiv:1207.1324}}].

\bibitem{Sun:2013cza}
S.~Sun, D.~B. Kaplan, and A.~E. Nelson, {\it {Little flavor: A model of
  weak-scale flavor physics}},  {\em Phys.Rev.} {\bf D87} (2013) 125036,
  [\href{http://xxx.lanl.gov/abs/1303.1811}{{\tt arXiv:1303.1811}}].

\bibitem{Buras:2003td}
A.~J. Buras, {\it {Relations between $\Delta M_{s,d}$ and $B_{s,d} \to \mu^+
  \mu^-$ in models with minimal flavour violation}},  {\em Phys. Lett.} {\bf
  B566} (2003) 115--119, [\href{http://xxx.lanl.gov/abs/hep-ph/0303060}{{\tt
  hep-ph/0303060}}].

\bibitem{Altmannshofer:2013foa}
W.~Altmannshofer and D.~M. Straub, {\it {New physics in $B \to
  K^*{\mu}{\mu}$?}},  \href{http://xxx.lanl.gov/abs/1308.1501}{{\tt
  arXiv:1308.1501}}.

\bibitem{Gauld:2013qba}
R.~Gauld, F.~Goertz, and U.~Haisch, {\it {On minimal $Z'$ explanations of the
  $B\to K^*\mu^+\mu^-$ anomaly}},
  \href{http://xxx.lanl.gov/abs/1308.1959}{{\tt arXiv:1308.1959}}.

\bibitem{Aaij:2013aka}
{\bf LHCb collaboration} Collaboration, R.~Aaij {\em et.~al.}, {\it
  {Measurement of the $B^0_s \to \mu^+ \mu^-$ branching fraction and search for
  $B^0 \to \mu^+ \mu^-$ decays at the LHCb experiment}},
  \href{http://xxx.lanl.gov/abs/1307.5024}{{\tt arXiv:1307.5024}}.

\bibitem{Chatrchyan:2013bka}
{\bf CMS Collaboration} Collaboration, S.~Chatrchyan {\em et.~al.}, {\it
  {Measurement of the $B_s \to \mu\mu$ branching fraction and search for $B_0
  \to \mu\mu$ with the CMS Experiment}},
  \href{http://xxx.lanl.gov/abs/1307.5025}{{\tt arXiv:1307.5025}}.

\bibitem{CMS-PAS-BPH-13-007}
{\it Combination of results on the rare decays $b\to \mu^+\mu^-$ from the cms
  and lhcb experiments},  Tech. Rep. CMS-PAS-BPH-13-007, CERN, Geneva, 2013.

\bibitem{Aaij:2013iag}
{\bf LHCb Collaboration} Collaboration, R.~Aaij {\em et.~al.}, {\it
  {Differential branching fraction and angular analysis of the decay $B^{0} \to
  K^{*0} \mu^{+}\mu^{-}$}},  \href{http://xxx.lanl.gov/abs/1304.6325}{{\tt
  arXiv:1304.6325}}.

\bibitem{Aaij:2013qta}
{\bf LHCb collaboration} Collaboration, R.~Aaij {\em et.~al.}, {\it
  {Measurement of form-factor independent observables in the decay $B^{0} \to
  K^{*0} \mu^+ \mu^-$}},  \href{http://xxx.lanl.gov/abs/1308.1707}{{\tt
  arXiv:1308.1707}}.

\bibitem{Chatrchyan:2013cda}
{\bf CMS Collaboration} Collaboration, S.~Chatrchyan {\em et.~al.}, {\it
  {Angular analysis and branching fraction measurement of the decay $B^0 \to
  K^* \mu^+\mu^-$}},  \href{http://xxx.lanl.gov/abs/1308.3409}{{\tt
  arXiv:1308.3409}}.

\bibitem{Descotes-Genon:2013wba}
S.~Descotes-Genon, J.~Matias, and J.~Virto, {\it {Understanding the $B\to
  K^*\mu^+\mu^-$ Anomaly}},  {\em Phys. Rev. D 88,} {\bf 074002} (2013)
  [\href{http://xxx.lanl.gov/abs/1307.5683}{{\tt arXiv:1307.5683}}].

\bibitem{Khodjamirian:2010vf}
A.~Khodjamirian, T.~Mannel, A.~Pivovarov, and Y.-M. Wang, {\it {Charm-loop
  effect in $B \to K^{(*)} \ell^{+} \ell^{-}$ and $B\to K^*\gamma$}},  {\em
  JHEP} {\bf 1009} (2010) 089, [\href{http://xxx.lanl.gov/abs/1006.4945}{{\tt
  arXiv:1006.4945}}].

\bibitem{Beylich:2011aq}
M.~Beylich, G.~Buchalla, and T.~Feldmann, {\it {Theory of $B\to K^{(*)}l^+l^-$
  decays at high $q^2$: OPE and quark-hadron duality}},  {\em Eur.Phys.J.} {\bf
  C71} (2011) 1635, [\href{http://xxx.lanl.gov/abs/1101.5118}{{\tt
  arXiv:1101.5118}}].

\bibitem{Matias:2012qz}
J.~Matias, {\it {On the S-wave pollution of $B\to K* l^+l-$ observables}},
  {\em Phys.Rev.} {\bf D86} (2012) 094024,
  [\href{http://xxx.lanl.gov/abs/1209.1525}{{\tt arXiv:1209.1525}}].

\bibitem{Jager:2012uw}
S.~Jager and J.~M. Camalich, {\it {On $B\to V \ell \ell$ at small dilepton
  invariant mass, power corrections, and new physics}},  {\em JHEP} {\bf 1305}
  (2013) 043, [\href{http://xxx.lanl.gov/abs/1212.2263}{{\tt
  arXiv:1212.2263}}].

\bibitem{Guadagnoli:2013mru}
D.~Guadagnoli and G.~Isidori, {\it {BR($B_s\to \mu^+\mu^-$) as an electroweak
  precision test}},  \href{http://xxx.lanl.gov/abs/1302.3909}{{\tt
  arXiv:1302.3909}}.

\bibitem{Buras:2012fs}
A.~J. Buras and J.~Girrbach, {\it {Complete NLO QCD Corrections for Tree Level
  Delta F = 2 FCNC Processes}},  {\em JHEP} {\bf 1203} (2012) 052,
  [\href{http://xxx.lanl.gov/abs/1201.1302}{{\tt arXiv:1201.1302}}].

\bibitem{Buras:1990fn}
A.~J. Buras, M.~Jamin, and P.~H. Weisz, {\it {Leading and next-to-leading QCD
  corrections to $\varepsilon$ parameter and $B^0-\bar{B}^0$ mixing in the
  presence of a heavy top quark}},  {\em Nucl. Phys.} {\bf B347} (1990)
  491--536.

\bibitem{Bobeth:2003at}
C.~Bobeth, P.~Gambino, M.~Gorbahn, and U.~Haisch, {\it {Complete NNLO QCD
  analysis of $\bar B \to X_s \ell^+ \ell^-$ and higher order electroweak
  effects}},  {\em JHEP} {\bf 04} (2004) 071,
  [\href{http://xxx.lanl.gov/abs/hep-ph/0312090}{{\tt hep-ph/0312090}}].

\bibitem{Huber:2005ig}
T.~Huber, E.~Lunghi, M.~Misiak, and D.~Wyler, {\it {Electromagnetic logarithms
  in $\bar B\to X(s) l^+ l^-$}},  {\em Nucl.Phys.} {\bf B740} (2006) 105--137,
  [\href{http://xxx.lanl.gov/abs/hep-ph/0512066}{{\tt hep-ph/0512066}}].

\bibitem{Buras:1994dj}
A.~J. Buras and M.~Munz, {\it Effective hamiltonian for $b \to x_s e^+ e^-$
  beyond leading logarithms in the ndr and hv schemes},  {\em Phys. Rev.} {\bf
  D52} (1995) 186--195, [\href{http://xxx.lanl.gov/abs/hep-ph/9501281}{{\tt
  hep-ph/9501281}}].

\bibitem{Bobeth:1999mk}
C.~Bobeth, M.~Misiak, and J.~Urban, {\it {Photonic penguins at two loops and
  $m_t$-dependence of $BR(B\to X_s \ell^+ \ell^-)$}},  {\em Nucl. Phys.} {\bf
  B574} (2000) 291--330, [\href{http://xxx.lanl.gov/abs/hep-ph/9910220}{{\tt
  hep-ph/9910220}}].

\bibitem{Bobeth:1999ww}
C.~Bobeth, M.~Misiak, and J.~Urban, {\it {Matching conditions for $b\to s
  \gamma$ and $b\to s g$ in extensions of the standard model}},  {\em
  Nucl.Phys.} {\bf B567} (2000) 153--185,
  [\href{http://xxx.lanl.gov/abs/hep-ph/9904413}{{\tt hep-ph/9904413}}].

\bibitem{Buras:2011zb}
A.~J. Buras, L.~Merlo, and E.~Stamou, {\it {The Impact of Flavour Changing
  Neutral Gauge Bosons on $\bar{B}\to X_s \gamma$}},  {\em JHEP} {\bf 1108}
  (2011) 124, [\href{http://xxx.lanl.gov/abs/1105.5146}{{\tt
  arXiv:1105.5146}}].

\bibitem{deBruyn:2012wk}
K.~De~Bruyn, R.~Fleischer, R.~Knegjens, P.~Koppenburg, M.~Merk, {\em et.~al.},
  {\it {Probing New Physics via the $B^0_s\to \mu^+\mu^-$ Effective Lifetime}},
   {\em Phys.Rev.Lett.} {\bf 109} (2012) 041801,
  [\href{http://xxx.lanl.gov/abs/1204.1737}{{\tt arXiv:1204.1737}}].

\bibitem{Fleischer:2012fy}
R.~Fleischer, {\it {On Branching Ratios of $B_s$ Decays and the Search for New
  Physics in $B^0_s\to \mu^+\mu^-$}},  {\em Nucl.Phys.Proc.Suppl.} {\bf
  241-242} (2013) 135--140, [\href{http://xxx.lanl.gov/abs/1208.2843}{{\tt
  arXiv:1208.2843}}].

\bibitem{Buras:2012ru}
A.~J. Buras, J.~Girrbach, D.~Guadagnoli, and G.~Isidori, {\it {On the Standard
  Model prediction for BR($B_{s,d} \to \mu^+\mu^-)$}},  {\em Eur.Phys.J.} {\bf
  C72} (2012) 2172, [\href{http://xxx.lanl.gov/abs/1208.0934}{{\tt
  arXiv:1208.0934}}].

\bibitem{Beaujean:2012uj}
F.~Beaujean, C.~Bobeth, D.~van Dyk, and C.~Wacker, {\it {Bayesian Fit of
  Exclusive $b \to s \bar\ell\ell$ Decays: The Standard Model Operator Basis}},
   {\em JHEP} {\bf 1208} (2012) 030,
  [\href{http://xxx.lanl.gov/abs/1205.1838}{{\tt arXiv:1205.1838}}].

\bibitem{Bobeth:2012vn}
C.~Bobeth, G.~Hiller, and D.~van Dyk, {\it {General Analysis of $\bar{B} \to
  \bar{K}^{(*)}\ell^+ \ell^-$ Decays at Low Recoil}},  {\em Phys.Rev.} {\bf
  D87} (2013) 034016, [\href{http://xxx.lanl.gov/abs/1212.2321}{{\tt
  arXiv:1212.2321}}].

\bibitem{Beaujean:2013soa}
F.~Beaujean, C.~Bobeth, and D.~van Dyk, {\it {Comprehensive Bayesian Analysis
  of Rare (Semi)leptonic and Radiative B Decays}},
  \href{http://xxx.lanl.gov/abs/1310.2478}{{\tt arXiv:1310.2478}}.

\bibitem{Horgan:2013pva}
R.~R. Horgan, Z.~Liu, S.~Meinel, and M.~Wingate, {\it {Calculation of $B^0 \to
  K^{*0} \mu^+ \mu^-$ and $B_s^0 \to \phi \mu^+ \mu^-$ observables using form
  factors from lattice QCD}},  \href{http://xxx.lanl.gov/abs/1310.3887}{{\tt
  arXiv:1310.3887}}.

\bibitem{Colangelo:2010et}
G.~Colangelo, S.~Durr, A.~Juttner, L.~Lellouch, H.~Leutwyler, {\em et.~al.},
  {\it {Review of lattice results concerning low energy particle physics}},
  {\em Eur.Phys.J.} {\bf C71} (2011) 1695,
  [\href{http://xxx.lanl.gov/abs/1011.4408}{{\tt arXiv:1011.4408}}].

\bibitem{Hurth:2008jc}
T.~Hurth, G.~Isidori, J.~F. Kamenik, and F.~Mescia, {\it {Constraints on New
  Physics in MFV models: A Model-independent analysis of $\Delta F = $1
  processes}},  {\em Nucl. Phys.} {\bf B808} (2009) 326--346,
  [\href{http://xxx.lanl.gov/abs/0807.5039}{{\tt arXiv:0807.5039}}].

\bibitem{Carrasco:2013zta}
N.~Carrasco, M.~Ciuchini, P.~Dimopoulos, R.~Frezzotti, V.~Gimenez, {\em
  et.~al.}, {\it {B-physics from Nf=2 tmQCD: the Standard Model and beyond}},
  \href{http://xxx.lanl.gov/abs/1308.1851}{{\tt arXiv:1308.1851}}.

\bibitem{Laiho:2009eu}
J.~Laiho, E.~Lunghi, and R.~S. Van~de Water, {\it {Lattice QCD inputs to the
  CKM unitarity triangle analysis}},  {\em Phys. Rev.} {\bf D81} (2010) 034503,
  [\href{http://xxx.lanl.gov/abs/0910.2928}{{\tt arXiv:0910.2928}}]. Updates
  available on {\tt http://latticeaverages.org/}.

\bibitem{Buras:2013raa}
A.~J. Buras and J.~Girrbach, {\it {Stringent Tests of Constrained Minimal
  Flavour Violation through $\Delta F=2$ Transitions}},  {\em The European
  Physical Journal C} {\bf 9} (73) 2013,
  [\href{http://xxx.lanl.gov/abs/1304.6835}{{\tt arXiv:1304.6835}}].

\bibitem{Marciano:1987ja}
W.~Marciano and A.~Sirlin, {\it {Constraint on additional neutral gauge bosons
  from electroweak radiative corrections}},  {\em Phys.Rev.} {\bf D35} (1987)
  1672--1676.

\bibitem{Altmannshofer:2009ma}
W.~Altmannshofer, A.~J. Buras, D.~M. Straub, and M.~Wick, {\it {New strategies
  for New Physics search in $B \to K^{*} \nu \bar{\nu}$, $B \to K \nu
  \bar{\nu}$ and $B \to X_{s} \nu \bar{\nu}$ decays}},  {\em JHEP} {\bf 04}
  (2009) 022, [\href{http://xxx.lanl.gov/abs/0902.0160}{{\tt
  arXiv:0902.0160}}].

\bibitem{Amhis:2012bh}
{\bf Heavy Flavor Averaging Group} Collaboration, Y.~Amhis {\em et.~al.}, {\it
  {Averages of B-Hadron, C-Hadron, and tau-lepton properties as of early
  2012}},  \href{http://xxx.lanl.gov/abs/1207.1158}{{\tt arXiv:1207.1158}}.

\bibitem{Aaij:2013oba}
{\bf LHCb collaboration} Collaboration, R.~Aaij {\em et.~al.}, {\it
  {Measurement of $CP$ violation and the $B_s^0$ meson decay width difference
  with $B_s^0 \to J/\psi K^+K^-$ and $B_s^0\to J/\psi\pi^+\pi^-$ decays}},
  \href{http://xxx.lanl.gov/abs/1304.2600}{{\tt arXiv:1304.2600}}.

\bibitem{Bona:2006sa}
{\bf UTfit} Collaboration, M.~Bona {\em et.~al.}, {\it {The UTfit collaboration
  report on the unitarity triangle beyond the standard model: Spring 2006}},
  {\em Phys. Rev. Lett.} {\bf 97} (2006) 151803,
  [\href{http://xxx.lanl.gov/abs/hep-ph/0605213}{{\tt hep-ph/0605213}}].
  {Updates available on \texttt{http://www.utfit.org}.}

\bibitem{Straub:2013uoa}
D.~M. Straub, {\it {Constraints on new physics from rare (semi-)leptonic B
  decays}},  \href{http://xxx.lanl.gov/abs/1305.5704}{{\tt arXiv:1305.5704}}.

\bibitem{Altmannshofer:2013oia}
W.~Altmannshofer, {\it {The $B_s \to \mu^+\mu^-$ and $B_d \to\mu^+\mu^-$
  Decays: Standard Model and Beyond}},
  \href{http://xxx.lanl.gov/abs/1306.0022}{{\tt arXiv:1306.0022}}.

\bibitem{Colangelo:1996ay}
P.~Colangelo, F.~De~Fazio, P.~Santorelli, and E.~Scrimieri, {\it {Rare $B \to
  K^{(*)} \nu\bar\nu$ decays at $B$ factories}},  {\em Phys.Lett.} {\bf B395}
  (1997) 339--344, [\href{http://xxx.lanl.gov/abs/hep-ph/9610297}{{\tt
  hep-ph/9610297}}].

\bibitem{Buchalla:2000sk}
G.~Buchalla, G.~Hiller, and G.~Isidori, {\it {Phenomenology of non-standard Z
  couplings in exclusive semileptonic $b\to s$ transitions}},  {\em Phys. Rev.}
  {\bf D63} (2001) 014015, [\href{http://xxx.lanl.gov/abs/hep-ph/0006136}{{\tt
  hep-ph/0006136}}].

\bibitem{Blanke:2009am}
M.~Blanke, A.~J. Buras, B.~Duling, S.~Recksiegel, and C.~Tarantino, {\it {FCNC
  Processes in the Littlest Higgs Model with T-Parity: a 2009 Look}},  {\em
  Acta Phys.Polon.} {\bf B41} (2010) 657--683,
  [\href{http://xxx.lanl.gov/abs/0906.5454}{{\tt arXiv:0906.5454}}].

\bibitem{Eberhardt:2012gv}
O.~Eberhardt, G.~Herbert, H.~Lacker, A.~Lenz, A.~Menzel, {\em et.~al.}, {\it
  {Impact of a Higgs boson at a mass of 126 GeV on the standard model with
  three and four fermion generations}},  {\em Phys.Rev.Lett.} {\bf 109} (2012)
  241802, [\href{http://xxx.lanl.gov/abs/1209.1101}{{\tt arXiv:1209.1101}}].

\bibitem{Buras:2010pi}
A.~J. Buras, B.~Duling, T.~Feldmann, T.~Heidsieck, C.~Promberger, {\em
  et.~al.}, {\it {Patterns of Flavour Violation in the Presence of a Fourth
  Generation of Quarks and Leptons}},  {\em JHEP} {\bf 1009} (2010) 106,
  [\href{http://xxx.lanl.gov/abs/1002.2126}{{\tt arXiv:1002.2126}}].

\bibitem{Hou:2013kia}
G.~W.~S. Hou, {\it {Enhanced $B_d\to \mu^+ \mu^-$ Decay: What if?}},
  \href{http://xxx.lanl.gov/abs/1307.2448}{{\tt arXiv:1307.2448}}.

\bibitem{Blanke:2008yr}
M.~Blanke, A.~J. Buras, B.~Duling, K.~Gemmler, and S.~Gori, {\it {Rare K and B
  Decays in a Warped Extra Dimension with Custodial Protection}},  {\em JHEP}
  {\bf 03} (2009) 108, [\href{http://xxx.lanl.gov/abs/0812.3803}{{\tt
  arXiv:0812.3803}}].

\bibitem{Buras:2013dea}
A.~J. Buras, F.~De~Fazio, and J.~Girrbach, {\it {331 models facing new $b\to s
  mu^+ mu^-$ data}},  \href{http://xxx.lanl.gov/abs/1311.6729}{{\tt
  arXiv:1311.6729}}.

\bibitem{Schael:2013ita}
{\bf ALEPH Collaboration, DELPHI Collaboration, L3 Collaboration, OPAL
  Collaboration, LEP Electroweak Working Group} Collaboration, S.~Schael {\em
  et.~al.}, {\it {Electroweak Measurements in Electron-Positron Collisions at
  W-Boson-Pair Energies at LEP}},
  \href{http://xxx.lanl.gov/abs/1302.3415}{{\tt arXiv:1302.3415}}.

\bibitem{Altmannshofer:2008dz}
W.~Altmannshofer, P.~Ball, A.~Bharucha, A.~J. Buras, D.~M. Straub, {\em
  et.~al.}, {\it {Symmetries and Asymmetries of $B \to K^{*} \mu^{+} \mu^{-}$
  Decays in the Standard Model and Beyond}},  {\em JHEP} {\bf 0901} (2009) 019,
  [\href{http://xxx.lanl.gov/abs/0811.1214}{{\tt arXiv:0811.1214}}].

\bibitem{Descotes-Genon:2013vna}
S.~Descotes-Genon, T.~Hurth, J.~Matias, and J.~Virto, {\it {Optimizing the
  basis of ${B} \to {K}^{*}\ell^+ \ell^-$ observables in the full kinematic
  range}},  {\em JHEP} {\bf 1305} (2013) 137,
  [\href{http://xxx.lanl.gov/abs/1303.5794}{{\tt arXiv:1303.5794}}].

\bibitem{Hambrock:2013zya}
C.~Hambrock, G.~Hiller, S.~Schacht, and R.~Zwicky, {\it {$B\to K^*$ Form
  Factors from Flavor Data to QCD and Back}},
  \href{http://xxx.lanl.gov/abs/1308.4379}{{\tt arXiv:1308.4379}}.

\bibitem{Bouchard:2013mia}
C.~Bouchard, G.~P. Lepage, C.~Monahan, H.~Na, and J.~Shigemitsu, {\it {Standard
  Model predictions for $B\to Kll$ with form factors from lattice QCD}},  {\em
  Phys. Rev. Lett. 111,} {\bf 162002} (2013)
  [\href{http://xxx.lanl.gov/abs/1306.0434}{{\tt arXiv:1306.0434}}].

\bibitem{Bouchard:2013eph}
C.~Bouchard, G.~P. Lepage, C.~Monahan, H.~Na, and J.~Shigemitsu, {\it {Rare
  decay $B\to Kll$ form factors from lattice QCD}},  {\em Phys. Rev. D 88,}
  {\bf 054509} (2013) 054509, [\href{http://xxx.lanl.gov/abs/1306.2384}{{\tt
  arXiv:1306.2384}}].

\end{thebibliography}\endgroup
\end{document}